\documentclass{aastex}

\begin{document}

\title{A Model for Abundances in Metal-Poor Stars}
\author{Y.-Z. Qian\altaffilmark{1} and G. J. Wasserburg\altaffilmark{2}}
\altaffiltext{1}{School of Physics and Astronomy, University of
Minnesota, Minneapolis, MN 55455; qian@physics.umn.edu.}
\altaffiltext{2}{The Lunatic Asylum, Division of Geological and
Planetary Sciences, California Institute of Technology, Pasadena,
CA 91125.}

\begin{abstract}
A model is presented that seeks to quantitatively explain the
stellar abundances of $r$-process elements and other elements
associated with the $r$-process sites. It is argued that the
abundances of all these elements in stars with $-3\lesssim {\rm
[Fe/H]}<-1$ can be explained by the contributions of three
sources. The sources are: the first generations of very massive
$(\gtrsim 100\,M_{\odot})$ stars that are formed from Big Bang
debris and are distinct from SNII, and two types of SNII, the $H$
and $L$ events, which can occur only at [Fe/H]~$\gtrsim -3$. The
$H$ events are of high frequency and produce dominantly heavy
($A>130$) $r$-elements but no Fe (presumably leaving behind black
holes). The $L$ events are of low frequency and produce Fe and
dominantly light ($A\lesssim 130$) $r$-elements (essentially none
above Ba). By using the observed abundances in two
ultra-metal-poor stars and the solar $r$-abundances, the initial
or prompt inventory of elements produced by the first generations
of very massive stars and the yields of $H$ and $L$ events can be
determined. The abundances of a large number of elements in a
star can then be calculated from the model by using only the
observed Eu and Fe abundances. To match the model results and the
observational data for stars with $-3<{\rm [Fe/H]}<-1$ requires
that the solar $r$-abundances for Sr, Y, Zr, and Ba must be
significantly increased from the standard values. No such changes
appear to be required for all other elements. If the changes in
the solar $r$-abundances for Sr, Y, Zr, and Ba are not permitted,
the model fails at $-3<{\rm [Fe/H]}<-1$ but still works at
[Fe/H]~$\approx -3$ for these four elements. By using the
corrected solar $r$-abundances for these elements, good agreement
is obtained between the model results and data over the range
$-3<{\rm [Fe/H]}<-1$. No evidence of $s$-process contributions is
found in this region, but all the observational data in this
region now show regular increases of Ba/Eu above the 
standard solar $r$-process value.
Whether the solar $r$-components of Sr, Y,
Zr, and Ba used here to obtain a fit to the stellar data can be
reconciled with those obtained from solar abundances by
subtracting the $s$-components calculated from models 
is not clear.
\end{abstract}


\section{Introduction}
\label{intro} 
We present a phenomenological model that seeks to
explain abundances of a large number of elements in
ultra-metal-poor (UMP) stars with [Fe/H]~$\approx -3$ and extend
this model to metal-poor (MP) stars with $-3<{\rm [Fe/H]}<-1$.
Recently Burris et al. (2000) reported abundances of heavy
elements, both above and below Ba, in 70 Galactic halo stars with
a wide range in [Fe/H]. This new work and previous observations by
e.g., McWilliam et al. (1995) and Ryan, Norris, \& Beers (1996)
stimulated us to address the problem of chemical evolution in the
early Galaxy. Recent studies have discussed Galactic chemical
evolution of H to Zn (Timmes, Woosley, \& Weaver 1995) and of
heavy elements from Ba to Eu with both $s$-process and $r$-process
contributions (Travaglio et al. 1999). These studies have shown
general trends for abundance ratios of different elements to Fe as
a function of [Fe/H] that converge on solar values. Raiteri et
al. (1999) treated temporal evolution of asymptotic giant branch
(AGB) stars and rates of Type II supernovae (SNII) and simulated
Galactic Ba enrichment. Their results on Ba/Fe show a wide scatter
at low [Fe/H] as observed. This was attributed to local
inhomogeneities in the interstellar medium (ISM). In the present
work, we will show that the abundances in UMP and MP stars follow
almost quantitatively from a set of simple rules.

The present study is purely phenomenological and confined to
$-3\lesssim{\rm [Fe/H]}<-1$ where Type Ia supernovae (SNIa, an Fe
source) and low mass AGB stars (the dominant $s$-process source)
would not contribute significantly to the ISM. As pointed out by
Truran (1981), abundances of neutron-capture elements (e.g., Ba)
at low [Fe/H] are dominated by $r$-process contributions from SNII.
We focus on the
most extensive and self-consistent data set so far available, that
of Burris et al. (2000), but treat data sets of Magain (1989),
Gratton \& Sneden (1994), and Johnson \& Bolte (2001) as well. We
will present the abundances resulting from production by the first
generations of stars formed after the Big Bang and yields for two
hypothesized types of SNII. These results will be inferred from
the observational data and not from theoretical models for the
first generations of stars or for $r$-process production by SNII.
The $r$-process abundances ($r$-abundances) calculated from our
model for UMP and MP stars will be compared with the observational
data. More specifically, the $r$-abundance of an element that
represents the total abundance of all $r$-process isotopes of the
element is calculated for comparison with observations as there
are no data on isotopic $r$-abundances in stars other than the
sun. When referring to $r$-process nuclei, we consider both
those that are the direct products of a rapid neutron capture
scenario and those that may be produced by other processes closely
related to the $r$-process in the same SNII event. The latter would
include many nuclei with atomic masses of $A\sim 90$ that would be
made in the $\alpha$-process (Woosley \& Hoffman 1992). If a
sufficient neutron abundance existed at the end of the
$\alpha$-process, the nuclei with $A\sim 90$ would become the
seed nuclei to capture the neutrons during the subsequent
$r$-process. Otherwise, these nuclei would experience no further
processing and be a part of the SNII ejecta. The $\alpha$-process
could be the dominant source of Sr, Y and Zr.

\subsection{The Basis of the Phenomenological Model}
\label{model1}
It is assumed here that SNII are the source of $r$-process nuclei
($r$-nuclei). The abundances of the $r$-nuclei $^{182}$Hf and
$^{129}$I in the early solar system are well established. These
data require that the SNII responsible for $^{182}$Hf not produce
any significant amounts of $^{129}$I [Wasserburg, Busso, \&
Gallino 1996 (WBG)]. The frequency of this type of SNII was argued
to be $\sim (10^7\ {\rm yr})^{-1}$ for a standard reference mass
of hydrogen (see \S\ref{model2}). The nuclide $^{129}$I was attributed
to a different type of SNII with a frequency of $\sim (10^8\ {\rm
yr})^{-1}$. These hypothetical SNII types are called $H$ and $L$
events for the ``high'' and ``low'' frequency types, respectively. This
approach assigns the dominant production of nuclei with
$A>130$ (``heavy'' $r$-nuclei) to the $H$ events and the
nuclei with $A\lesssim 130$ (``light'' $r$-nuclei) to the $L$ events.
Qian, Vogel, \& Wasserburg (1998) developed a simple
$r$-process model involving neutrinos to examine the different
conditions in $H$ and $L$ events that are needed to provide a
split between the production of heavy and light $r$-nuclei at
$A\sim 130$. Due to their much higher frequency, the $H$ events
were expected to be likely the first SNII to inject $r$-nuclei
into the ISM (WBG). The effects of such
``first generation" SNII should be
apparent in the abundances of $r$-elements in UMP stars.  Exquisite
observations by Sneden et al. (1996, 2000) and Westin et al.
(2000) have established that abundance ratios of many other heavy
$r$-elements relative to Eu in UMP stars are remarkably constant
and close to solar $r$-process values. This demonstrates that
abundances of heavy $r$-elements in UMP stars appear to reflect
production by pure $H$ events. Sneden et al. (2000) also found
that the abundances of light $r$-elements such as Pd, Ag, and Cd
in the UMP star CS 22892-052 were low relative to the solar
$r$-abundance pattern that is
translated to pass through the Eu data. This
provides additional evidence that there should be at least two
distinct types of SNII sources for the $r$-nuclei.

The available data on UMP stars (e.g., McWilliam et al. 1995;
McWilliam 1998; Westin et al. 2000; Sneden et al. 2000; Burris et
al. 2000) show that there is a wide range in the abundances of the
heavy $r$-elements Eu and Ba ($\sim 2$ dex) for stars with
$-3\lesssim{\rm [Fe/H]}<-2.5$. This clearly indicates that heavy
$r$-elements, including the chronometer Th, are produced without
any significant coproduction of Fe, thus requiring that heavy
$r$-elements and Fe be produced in different types of SNII. There
is little Ba in stars with $-4\lesssim {\rm [Fe/H]}<-3$ (McWilliam
et al. 1995; McWilliam 1998). Wasserburg \& Qian [2000 (WQ)]
inferred that the large dispersion in Eu and Ba abundances at
[Fe/H]$\sim -3$ was caused by $H$ events adding heavy $r$-elements
but no Fe to a pre-existing inventory of elements in the ISM. This
pre-existing inventory of Fe and associated elements was called
the initial or prompt (hereafter $P$) inventory and attributed to
the first generations of very massive ($\gtrsim 100\,M_\odot$)
stars formed from Big Bang debris. It was further inferred that
normal stars with masses of $\sim(1$--$60)\,M_\odot$ could only
form in a medium with sufficient metals to permit cooling during
the collapse of a gas cloud. Normal stars with masses of
$\sim(10$--$60)\,M_\odot$ would later become SNII. Further
evidence for the decoupling of heavy $r$-elements from Fe has been
provided by the observations of the U-rich star CS 31082-001
(Cayrel et al. 2001). This star with [Fe/H]~$=-2.9$ has extremely
abundant Os, Ir, Th, and U, sharply exhibiting the effect that
little Fe is coproduced with the heavy $r$-elements as argued by
Qian \& Wasserburg [2001b (QW)]. Figure \ref{ba} shows
the available Ba data over $-4\lesssim {\rm [Fe/H]}\lesssim -1$
and includes the Ba abundance inferred by QW for CS 31082-001
based on the three-component ($P$, $H$, and $L$) model for 
abundances in UMP and MP stars. This inferred abundance (upper cross)
is within $\approx 0.3$~dex of the preliminary Ba data on 
CS 31082-001 (lower cross)
reported by Hill et al. (2001). As discussed above, the large
scatter in Ba abundance at $-3\lesssim {\rm [Fe/H]}<-2.5$ shown in
Figure \ref{ba} is attributed to the rapid occurrence of $H$
events shortly after the onset of normal star formation.

The scenario adopted here for the evolution of $r$-process related
elements over Galactic history is schematically shown in Table
\ref{epoch}. Some time after the Big Bang, baryonic matter
condenses in dark matter potential wells. Only very massive stars
can form from Big Bang debris. These stars explode and contribute
newly-synthesized material. The number of such events is not known
but many such disrupting events may be required to give
[Fe/H]~$\approx -3$ in an average parcel of the
ISM or IGM (intergalactic
medium). The ISM or IGM now has a changed chemical composition
compared with Big Bang debris. When some of this material cools,
there is condensation of gas to form a more or less normal stellar
population that includes the progenitors for SNII $H$ and $L$
events and lower mass stars. This state of condensation can be
achieved when the abundance of all metals in the ISM or IGM
reaches or surpasses some critical value. A nominal value of
[Fe/H]~$\approx -3$ for the critical ``metallicity'' was inferred
by WQ from the observational data on UMP stars. This appears to be
supported by the recent results of Bromm et al. (2001) who have
simulated the collapse and fragmentation of gas clouds with
different metallicities. The nominal value [Fe/H]~$\approx -3$
will be used below for the $P$-inventory of Fe although the onset
of ``normal'' astration may correspond to a possible range of
$-4<{\rm [Fe/H]}< -2.7$. The first SNII to contribute $r$-elements
to the ISM are the most frequent and most probable $H$ events. On
a timescale of $\gg 10^8$~yr, the relative contributions of $H$
and $L$ events should approach well-defined values that are 
determined by the average frequency ratio of
$\sim 10:1$ and the yields of these events. From
then on the evolution of $r$-elements and Fe would progress as if
there were only one type of SNII with small statistical deviations
(see the evolution of Ba abundance relative to Fe at
[Fe/H]~$>-2.5$ indicated by the dot-dashed line in Figure
\ref{ba}). At [Fe/H]~$\sim -1$, SNIa begin to contribute and
dominate the Fe production, as inferred from the O data (e.g.,
Timmes et al. 1995; Qian \& Wasserburg 2001a).

This approach sharply separates heavy element production in the
earliest epochs [(2) and (3) in Table \ref{epoch}] from later 
production by
SNII [(6) and (7)] and from even later production by SNIa and
contributions of evolved normal low-mass stars. The numbers of $H$
and especially $L$ events through epochs (6) and (7) are not very
large (see \S\ref{nhl}) so that a discrete model is required.

For simplicity it is assumed that each $H$ or $L$ event has a
fixed relative abundance pattern for the elements produced in each
case and that the yields are constant for each $H$ or $L$ event.
The average frequencies of these events are also assumed to be
constant for a standard reference mass of hydrogen. The main
results of the model only depend on the assumption of fixed yield
patterns for $H$ and $L$ events (see \S\S\ref{model2}--\ref{nhl}).
The key parameters of the model that must be determined by
consideration of the observational data are the yields of $H$ and
$L$ events for pertinent  elements and the $P$-inventory
composition that defines the baseline  to which the contributions
of $H$ and $L$ events are added. As we are  concerned with the
addition of elements to the ISM or IGM whose  hydrogen mass
fraction is essentially not altered during Galactic chemical
evolution, we will typically use a representation where the
abundances are given relative to hydrogen. This is the standard
spectroscopic notation $\log \epsilon({\rm E})\equiv\log({\rm
E/H})+12$ for  element E where E is also used to represent the
number of E atoms. The common notation [E/Fe] complicates the
discussion and is avoided here as ranges in [E/Fe] may result from
addition of either element E or Fe.

\subsection{Time-Dependent Evolution and Mixing}
\label{model2}
Consider a homogeneous system of gas with a time-dependent total number
H of hydrogen atoms. The rate of change in the total number E of E atoms
in this gas may be written as
\begin{equation}
\label{dedt}
\frac{d{\rm E}}{dt} = \sum_i P_{{\rm E},i} +
\left(\frac{\rm E}{\rm H}\right)\frac{d{\rm H}}{dt},
\end{equation}
where $\sum_i P_{{\rm E},i}$ is the total rate for injection of
element E into the gas after production by its sources, and the
last term accounts for the removal of matter from the gas
by either astration or fragmentation ($d{\rm H}/dt<0$ for
both cases). The above equation may be rewritten as
\begin{equation}
\label{dehdt}
\frac{d({\rm E/H})}{dt} = \sum_i P_{{\rm E},i}/{\rm H}.
\end{equation}
Thus, if SNII $H$ and $L$ events are the only sources under
consideration for element E and the SNII frequencies are
proportional to the amount of hydrogen in the gas, then equation
(\ref{dehdt}) reduces to the following for a homogeneous mass of
gas:
\begin{equation}
\label{eh}
({\rm E/H})=({\rm E/H})_P+n_H({\rm E/H})_H+n_L({\rm E/H})_L,
\end{equation}
or in the spectroscopic notation,
\begin{equation}
\label{eps}
10^{\log\epsilon({\rm E})}=10^{\log\epsilon_P({\rm E})}
+n_H\times 10^{\log\epsilon_H({\rm E})}
+n_L\times 10^{\log\epsilon_L({\rm E})}.
\end{equation}
In equation (\ref{eh}),  (E/H)$_P$ is the
$P$-inventory, (E/H)$_H$ or (E/H)$_L$ is the number of E atoms
produced per hydrogen atom in the gas phase by each $H$ or $L$
event for a standard reference mass of hydrogen that is
assumed here to
mix with the ejecta in each case, and $n_H$ or $n_L$ is the number
of $H$ or $L$ events that have occurred in this reference mass.

Now consider the mixture of two different gas systems (with
subscripts ``1'' and ``2'') that have different nucleosynthetic
histories. For fixed values of (E/H)$_P$, (E/H)$_H$, and
(E/H)$_L$, the abundance of element E in the mixture of two gas
masses can be written as
\begin{eqnarray}
\left(\frac{\rm E}{\rm H}\right)_{\rm mix} & = &
\frac{{\rm (E/H)}_1({\rm H})_1 + {\rm (E/H)}_2({\rm H})_2}
{({\rm H})_1 + ({\rm H})_2}\\  \nonumber
& = &({\rm E/H})_P+\tilde n_H({\rm E/H})_H+
\tilde n_L({\rm E/H})_L,
\end{eqnarray}
where
\begin{mathletters}
\begin{eqnarray}
\tilde{n}_H &=& \frac{(n_H)_1({\rm H})_1 + (n_H)_2({\rm H})_2}
{({\rm H})_1 + ({\rm H})_2},\\
\tilde{n}_L &=& \frac{(n_L)_1({\rm H})_1 + (n_L)_2({\rm H})_2}
{({\rm H})_1 + ({\rm H})_2}.
\end{eqnarray}
\end{mathletters}
It can be seen that a mixture of two gas systems with different
nucleosynthetic histories will result in a new system with
effective values of $\tilde{n}_H$ and $\tilde{n}_L$ corresponding
to the weighted averages of the contributing systems. As a result,
all that mixing will do is change the values of $n_H$ and $n_L$
but still produce abundance patterns congruent to those resulting
from $H$ and $L$ events. If the absolute yields of $H$ and $L$
events were to shift but the relative yields of one element to the
other be maintained in each type of event, then the only result of
this shift would again be to change $n_H$ and $n_L$ but the same
congruence of abundance patterns would be maintained. It follows
that a wide range of removal by astration and mixing of matter
with different chemical evolution lead to a result that is
indistinguishable from a simple history if the $P$-inventory
composition and the basic relative yield templates of $H$ and $L$
events are fixed as assumed in this model. Thus complex mixing and
transport models are not required to pursue the approach used here
but are effectively represented by the $n_H$ and $n_L$ values for
a star. If the model assumptions are correct, then they provide a
full basis for quantitatively determining the $r$-abundances in
UMP and MP stars.

The frequencies of $H$ and $L$ events inferred from the abundances
of $^{182}$Hf and $^{129}$I in the early solar system, $\sim
(10^7\ {\rm yr})^{-1}$ and $\sim (10^8\ {\rm yr})^{-1}$,
respectively, correspond to a reference mass of hydrogen of
$M_{\rm H}^{\rm ref}\sim 3\times 10^4\,M_\odot$ for a total SNII rate
of $\sim (30\ {\rm yr})^{-1}$ and a total gas mass of
$\sim 10^{10}\,M_\odot$ in the present Galaxy
(Qian \& Wasserburg 2001a). This reference mass is also the total amount
of ISM typically swept by an SNII remnant (e.g., Thornton et al. 1998).
We take this as the standard reference mass of hydrogen  to mix with the
nucleosynthetic products of an SNII $H$ or $L$ event. We will use
equation (\ref{eps}) with fixed parameters $\log\epsilon_P({\rm E})$,
$\log\epsilon_H({\rm E})$, and  $\log\epsilon_L({\rm E})$ to discuss
abundances in UMP and MP stars.  All quantities in this equation will be
calculated for the above reference mass of hydrogen. The absolute yield of
element E in units of mass for an $H$ or $L$ event, $Y_H({\rm E})$ or
$Y_L({\rm E})$, can be estimated from the relationships:
\begin{mathletters}
\begin{eqnarray}
\left({{\rm E}\over {\rm H}}\right)_H &=&
{Y_H({\rm E})/\langle A_{\rm E}\rangle_H\over
M_{\rm H}^{\rm ref}},\\
\left({{\rm E}\over {\rm H}}\right)_L &=&
{Y_L({\rm E})/\langle A_{\rm E}\rangle_L\over M_{\rm H}^{\rm ref}},
\end{eqnarray}
\end{mathletters}
where $\langle A_{\rm E}\rangle_H$ or $\langle A_{\rm E}\rangle_L$
is the average atomic mass for the E isotopes produced in an $H$ or
$L$ event.

In discussing the stellar observations, we consider that the
observed stars formed from the immediate precursor ISM with the
homogenized $r$-abundances of that local region. These
$r$-abundances are not considered to have been altered by the
subsequent evolution of the star. If the surface of a star is
contaminated by some form of mass transfer from a companion (or
nearby) SNII, then the observed abundance pattern should represent
the products of that SNII (an $H$ or $L$ event), but the
abundances would be greatly enhanced from those with which the
star formed. For example, the extremely high Ba abundances in
CS 31082-001 (crosses linked by a line) and CS 22892-052 (open circle)
shown in Figure \ref{ba} may have resulted from surface contamination
by the SNII explosion (of the $H$ type) of their respective
binary companions (QW).
The issue of what level of $r$-process enrichment is so large
that an SNII companion (or neighbor) appears to be required has been
discussed by QW and will be considered
below (see \S\ref{syzbump}).

\subsection{Calculation of $n_H$ and $n_L$}
\label{nhl}
Given the $P$-inventory,
$\log\epsilon_P({\rm E})$,  and the yields of $H$ and $L$ events,
$\log\epsilon_H({\rm E})$ and $\log\epsilon_L({\rm E})$,
the abundance of element E in a star, $\log\epsilon({\rm E})$,
can be calculated from equation (\ref{eps}). The numbers of
contributing $H$ and $L$ events for the star, $n_H$ and $n_L$,
are required for this calculation. If a certain element is exclusively
produced by $H$ or $L$ events, then $n_H$ or $n_L$ can be
simply obtained from the observed abundance of this element
in the star by using the yield of an $H$ or $L$ event for this
element.

Data on Ba ($A\sim 135$) at $-4\lesssim[{\rm Fe/H}]<-3$
(McWilliam et al.
1995; McWilliam 1998) indicate that the $P$-inventory of heavy
$r$-elements is negligible (WQ; see Figure \ref{ba}). 
Observations also show that the
abundances of Ba and above in UMP stars very closely follow the
solar $r$-abundance pattern (Sneden et al. 1996, 2000; Westin et
al. 2000). We consider that $H$ events are the exclusive source
for the heavy $r$-elements above Ba (possibly including Ba)
with a yield pattern
identical to the corresponding part of the solar  $r$-abundance
pattern (see \S\S\ref{phcomp} and \ref{heavy}). Then the $H$-yield of
any heavy $r$-element above Ba can be used to calculate the value
$n_H$ for a star. We use Eu ($A\sim 151$) for this purpose as it
is typically observed in UMP and MP stars and its solar inventory
is dominated by $r$-process contributions. Specifically, the solar
$r$-abundance of Eu, $\log\epsilon_{\odot,r}({\rm Eu})=0.52$
(Arlandini et al. 1999), was contributed by $n_H^\odot=10^3$ $H$
events for a frequency of $f_H=(10^7\ {\rm yr})^{-1}$ over a
period of $10^{10}$~yr prior to solar system formation (SSF; see
\S\ref{nhlsun}). Thus $\log\epsilon_H({\rm
Eu})=\log\epsilon_{\odot,r}({\rm Eu}) -\log n_H^\odot=-2.48$. The
value $n_H$ for a star is then:
\begin{equation}
\label{eu} n_H=10^{\log\epsilon({\rm Eu})-\log\epsilon_H({\rm
Eu})} = 10^{\log\epsilon({\rm Eu})+2.48}.
\end{equation}

As little or no Fe is produced in $H$ events (see \S\ref{model1}),
we consider that the Fe abundance at $-3<[{\rm Fe/H}]<-1$ results
from contributions of $L$ events being added to the
$P$-inventory:
\begin{equation}
\label{epsfe}
10^{\log\epsilon({\rm Fe})}=10^{\log\epsilon_P({\rm Fe})}
+n_L\times 10^{\log\epsilon_L({\rm Fe})}.
\end{equation}
As SNIa were the dominant Fe source at
[Fe/H]~$>-1$, only a fraction $\alpha_{\rm Fe}$ of the solar Fe
inventory $\log\epsilon_\odot({\rm Fe})=7.51$ (Anders \&
Grevesse 1989) was contributed by SNII $L$ events. For
$\alpha_{\rm Fe}=1/3$ (e.g., Timmes et al. 1995;
Qian \& Wasserburg 2001a)
and with $n_L^\odot=10^2$ contributing $L$ events
for a frequency of $f_L=(10^8\ {\rm yr})^{-1}$ over a period of
$10^{10}$~yr prior to SSF (see \S\ref{nhlsun}),
$\log\epsilon_L({\rm Fe})=\log\alpha_{\rm Fe}+
 \log\epsilon_\odot({\rm Fe})-\log n_L^\odot=5.03$
corresponding to [Fe/H]$_L=\log\epsilon_L({\rm Fe})-
\log\epsilon_\odot({\rm Fe})=-2.48$. We take $\log\epsilon_P({\rm
Fe})\approx 4.51$ corresponding to [Fe/H]$_P\approx -3$. The
$P$-inventory of Fe is overwhelmed by Fe addition from a few $L$
events. Thus the possible range in [Fe/H]$_P$ has little effect on
the determination of $n_L$ for MP stars. By using the above
$P$-inventory and $L$-yield of Fe, the value $n_L$ for a star can
be obtained from equation (\ref{epsfe}) or from its equivalent in
the [Fe/H] notation:
\begin{equation}
\label{fe}
10^{\rm [Fe/H]}=10^{{\rm [Fe/H]}_P}+
n_L\times 10^{{\rm [Fe/H]}_L}.
\end{equation}

In all our works, SNII $H$ and $L$ events are considered to
occur stochastically with frequencies of $f_H$ and $f_L$ in some
standard reference mass of hydrogen throughout the Galaxy over
Galactic history (WBG; WQ;
Qian \& Wasserburg 2000, 2001a, QW). At any
given time $t$ since the onset of SNII, the probability $P(n_H,t)$
for a standard parcel of the ISM to have had a number $n_H$ of $H$
events is given by the Poisson distribution:
\begin{equation}
\label{pnht}
P(n_H,t)={(f_Ht)^{n_H}\over n_H!}e^{-f_Ht} = {(\bar n_H)^{n_H}
\over n_H!}e^{-\bar n_H},
\end{equation}
where $\bar n_H=f_Ht$ is the average number of occurrences for $H$
events at time $t$. For $L$ events a similar expression applies:
\begin{equation}
\label{pnlt}
P(n_L,t)={(f_Lt)^{n_L}\over n_L!}e^{-f_Lt}={(\bar n_L)^{n_L}
\over n_L!}e^{-\bar n_L}.
\end{equation}
The variances of the distributions in equations (\ref{pnht}) and
(\ref{pnlt}) are $\bar n_H$ and $\bar n_L$, respectively. The square
root of the variance, $\sqrt{\bar n_H}$ or $\sqrt{\bar n_L}$, can be used
as a measure of the variation in $n_H$ or $n_L$ at time $t$:
$n_H\approx \bar n_H\pm\sqrt{\bar n_H}$ or
$n_L\approx \bar n_L\pm\sqrt{\bar n_L}$. At $t=10^9$~yr,
$n_H \approx 100\pm 10$ corresponding to
$\log\epsilon({\rm Eu})\approx -0.53$ to $-0.44$, while
$n_L\approx 10\pm 3$ corresponding to
[Fe/H]~$\approx -1.62$ to $-1.36$. For shorter
times, the dispersion in $n_L$ becomes much more prominent. For
example, at $t=3\times 10^8$~yr,
$n_H \approx 30\pm 5$ corresponding to
$\log\epsilon({\rm Eu})\approx -1.08$ to $-0.94$, while
$n_L\approx 3\pm 2$ corresponding to
[Fe/H]~$\approx -2.37$ to $-1.76$. The abundances of heavy
$r$-elements (e.g., Eu) are a better measure of time because of the higher
frequency of $H$ events, and hence much lower dispersion in $n_H$.
Thus it follows that there is a substantial dispersion in
[Fe/H] for a rather fixed abundance of any heavy $r$-element at
any given time in the range $10^8\lesssim t\lesssim 2\times10^9$~yr.
Conversely, a given [Fe/H] value may correspond to a significant range
of possible times. We consider that this effect causes the scatter in
abundances of heavy $r$-elements at $-3\lesssim{\rm [Fe/H]} < -1.0$
with the dispersion rapidly increasing at lower [Fe/H] . The dispersion
and mean trends for abundances of heavy and light $r$-elements in
UMP and MP stars have been discussed in some detail by Qian (2001)
based on the same approach.

As discussed in \S1.2, the complex history of mixing and transport
of elements in the ISM from which a star formed is represented by
$n_H$ and $n_L$ for the star. It is shown above that these numbers
can be calculated from the observed Eu and Fe abundances in the
star by using the $H$-yield of Eu and the $L$-yield of Fe.
Substituting the expressions for $n_H$
(eq. [\ref{eu}]) and $n_L$ (eq. [\ref{epsfe}] or [\ref{fe}]) into the
expression for the abundance of element E
(eq. [\ref{eh}] or [\ref{eps}]), we obtain
\begin{equation}
\label{eeufe}
({\rm E/H})=({\rm E/H})_P+({\rm E/Eu})_H({\rm Eu/H})
+({\rm E/Fe})_L[({\rm Fe/H})-({\rm Fe/H})_P].
\end{equation}
As the yield of element E relative to Eu for $H$ events, (E/Eu)$_H$,
or to Fe for $L$ events, (E/Fe)$_L$, is assumed to be fixed,
the observed Eu or Fe abundance in the star contains all the prior
history of mixing and
transport for the part of element E coproduced with Eu in $H$ events
or with Fe in $L$ events. Thus the fundamental problem in our
approach to understand abundances in UMP and MP stars is to
establish the $P$-inventory composition and the yield
templates of $H$ and $L$ events that underlie the observations.
These results will be presented in the form of $\log\epsilon_P({\rm E})$,
$\log\epsilon_H({\rm E})$, and $\log\epsilon_L({\rm E})$
corresponding to $\log\epsilon_P({\rm Fe})\approx 4.51$
([Fe/H]~$\approx -3$), $\log\epsilon_H({\rm Eu})=-2.48$, and
$\log\epsilon_L({\rm Fe})=5.03$ ([Fe/H]$_L=-2.48$).

\subsection{Values of $n_H^{\odot}$ and $n_L^{\odot}$}
\label{nhlsun} 
We assume here that the solar abundances are
representative of an average parcel of the
ISM at the time of SSF. We use the
values $n_H^{\odot}=10^3$ and $n_L^{\odot}=10^2$ to account for
the part of the solar inventory that was contributed by SNII over
a total uniform production time of $T_{\rm UP}\sim 10^{10}$~yr
prior to SSF. These values are directly related to the frequencies
of $H$ and $L$ events that were derived from consideration of
meteoritic data on the abundances of $^{182}$Hf (with a lifetime
$\bar\tau_{182}=1.3\times 10^7$~yr) and $^{129}$I
($\bar\tau_{129}=2.3\times 10^7$~yr) in the early solar system
(WBG). The meteoritic data require that replenishment of fresh
$^{182}$Hf in an average parcel of the
ISM must occur on a timescale
commensurate with its lifetime, which corresponds to a frequency
of $f_H\sim(10^7\ {\rm yr})^{-1}$ for $H$ events in a standard
reference mass of hydrogen. As discussed in \S\ref{model2}, this
frequency can be explained by considering the Galactic SNII rate
per unit mass of gas and the total amount of ISM typically swept
up by an SNII remnant. Although there may be a range of possible
values for $n_H^\odot=f_HT_{\rm UP}$, the observed abundances of
heavy $r$-elements in UMP and MP stars appear to be well
represented by the $H$-yields of these elements calculated by
using $n_H^\odot=10^3$. For example, the lowest Eu abundances
observed in such stars (e.g., McWilliam et al. 1995; Burris et al.
2000) are remarkably close to the Eu yield of a single $H$ event
[$\log\epsilon_H({\rm Eu}) =-2.48$] calculated from the solar
$r$-inventory of Eu for $n_H^\odot=10^3$. In any case, a change in
$n_H^\odot$ causes only a uniform shift in the $H$-yields of the
relevant elements as the yield template is assumed here to be
fixed. Such a shift will not affect the basic results presented
here (see \S\S\ref{model2}--\ref{nhl}).

The timescale for replenishment of fresh $^{129}$I in an average
parcel of the
ISM required by the meteoritic data is much longer than
its lifetime and corresponds to a frequency of $f_L\sim (10^8\
{\rm yr})^{-1}$ for $L$ events. A decrease in the replenishment
timescale (corresponding to an increase in $f_L$) would cause an
exponential increase in the abundance of $^{129}$I in the early
solar system (ESS).  Many meteoritic measurements give a rather
precise and very low value for this abundance
[$(^{129}$I/$^{127}$I)$_{\rm ESS}=10^{-4}$; e.g., 
Brazzle et al. 1999],
severely limiting any possible increase in $f_L$. Thus we focus on
the effect of decreasing $n_L^\odot=f_LT_{\rm UP}$ from $10^2$.
For $n_L^\odot=10^2$ the $L$-yield of Fe is [Fe/H]$_L=-2.48$. A
decrease in $n_L^\odot$ by a factor of 2 gives [Fe/H]$_L=-2.18$,
which means that there would be a lack of stars between
[Fe/H]~$\approx -3$ and $-2.18$. This is incompatible with the
observations that show a continuous population at [Fe/H]~$\gtrsim
-2.5$. In any case, a change in $n_L^\odot$ again causes only a
uniform shift in the $L$-yields of the relevant elements. Thus
changes in the values of $n_H^{\odot}$ and $n_L^{\odot}$ will not
affect the basic conclusions of this work. But if the changes are
large, they will violate the constraints indicated above. The
standard values $n_H^{\odot}=10^3$ and $n_L^{\odot}=10^2$ will be
used to derive the results in this work.

\subsection{The Solar $r$-Component}
\label{rsun} 
In this study we are mainly concerned with the
$r$-process. As discussed in \S\ref{model1}, there must be at
least two distinct types of $r$-process events. We associate both
of these with SNII, although they have not yet been proven to be
the $r$-process sites. It was argued that the other candidate
sites, such as neutron star mergers, would have difficulty
explaining the observed $r$-abundances in UMP and MP stars if they
were the principal source for $r$-nuclei (Qian 2000). Extensive
and very sophisticated parametric studies of high neutron density
environments have been carried out to gain insights into the
conditions that are required for the $r$-process (Kratz et al.
1993; Hoffman, Woosley, \& Qian 1997; Meyer \& Brown 1997;
Freiburghaus et al. 1999). These studies are usually directed
towards approximating the overall solar $r$-abundance pattern by
the $r$-process production in a prototypical environment. However,
there is as yet no sufficient guidance from these studies to
explain the distinct types of $r$-process events discussed in
\S\ref{model1} or to make reliable {\it a priori} predictions for
the templates of  $r$-process yields ($r$-yields) that can be used
in Galactic chemical  evolution models.

As also discussed in \S\ref{model1}, the first generations of 
very massive
stars formed from Big Bang debris produced a $P$-inventory of
Fe and associated elements but very little heavy $r$-nuclei. The
formation and evolution
of these stars and their nucleosynthetic products have
been a long-standing problem (e.g., Ezer \& Cameron 1971) and
recently received renewed attention (e.g., Bromm, Coppi, \& Larson
2000; Heger et al. 2000; Fryer et al. 2001;
Baraffe, Heger \& Woosley 2001; Bromm et al. 2001).
However, so far there is not sufficient information from 
theoretical models
to predict the composition of the $P$-inventory assigned  here
to the first generations of stars.

Our approach here is to determine some of the $P$-inventory
composition and the $r$-yields of SNII $H$ and $L$ events from the
observed abundances in several selected stars based on the
phenomenological model laid out in \S\S\ref{model1}--\ref{nhl}.
These results will be used in the model to calculate the
abundances for UMP and MP stars in general. The calculated
abundances will then be compared with observations. One of the
selected stars is the sun whose so-called ``cosmic" abundance
pattern (mostly derived from meteoritic studies) has served as the
key to understanding nucleosynthesis (Burbidge et al. 1957;
Cameron 1957). The part of the solar abundances required here is
the $r$-process component ($r$-component) that is obtained by
subtracting the $s$-process component ($s$-component) from the
total solar inventory.

One of the better understood processes of stellar nucleosynthesis
is the $s$-process (see  review by Busso, Gallino, \& Wasserburg
1999). The site of this process is well established
observationally to be thermally pulsing AGB stars. Theoretical
models for the $s$-process are based on the well (but not
completely) understood laws of stellar structure and evolution and
the nuclear reactions in the stellar interior that produce the
neutrons for slow capture by the pre-existing seed nuclei. The
nuclear physics involved both in neutron production and in
$s$-processing has been the subject of extensive theoretical and
experimental work. Many cross sections for important neutron
capture reactions have been reliably measured. The classical
analysis of the $s$-process relies on mainly the nuclear physics
input and adopts a suitable distribution of neutron exposures that
in principle can be calculated from stellar models. This approach is
particularly successful for the so-called main $s$-component
(see K\"appeler, Beer, \& Wisshak 1989 for a review). The
most recent $s$-process calculations of Arlandini et al. (1999)
have emphasized the dependences of  $s$-process yields on the
initial mass and metallicity of AGB stars. Their calculations were
embedded in very high quality stellar models with highly resolved
structure of thermal pulses. They showed that the classical main
$s$-component was very well reproduced by averaging the yields for
models of $1.5\,M_\odot$ and $3\,M_\odot$ AGB stars with 
$Z=0.5Z_\odot$
($Z$ is the total mass fraction of all elements above He and
$Z_\odot$ is the solar value).

However, there are still questions with regard to the
one-dimensional stellar models used in the new $s$-process
calculations, such as the conditions obtained at critical boundary
layers in AGB stars. One fundamental problem is the formation of
the $^{13}$C pocket that is the dominant source of neutrons [via
$^{13}{\rm C}(\alpha,n)^{16}{\rm O}]$ in the new stellar models.
By assuming that the mass of the $^{13}$C pocket is constant for
AGB stars with different metallicities, it was shown that the
neutron-to-seed ratio for the $s$-process increases with
decreasing metallicity, giving rise to preponderant production of
heavier elements by low metallicity AGB stars (Gallino et al.
1998). This result is very plausible. However, it appears to be in
conflict with the recent observations of LP 625-44 (Aoki et al.
2000). This star has [Fe/H]~$=-2.7$ and is enormously enriched in
heavy elements from Ba and above. The high enrichments in LP
625-44 were attributed to surface contamination 
by mass transfer from an evolved AGB companion with the same
metallicity [Fe/H]~$=-2.7$ (Aoki et al. 2000)
and do not reflect significant
$s$-process addition to the ISM from which the star formed. The
abundance pattern from Ba and above in this star matches that of
the main $s$-component calculated by Arlandini et al. (1999) from
models of AGB stars but with $Z=0.5Z_\odot$. As pointed out by
Aoki et al. (2000), this abundance pattern is not in accord with
models of low metallicity AGB stars that have very high
neutron-to-seed ratios: the observed Pb/Ba ratio is lower by two
orders of magnitude than the value predicted by Busso et al.
(1999) from stellar models with [Fe/H]~$=-2.7$. On the other hand,
the observed abundances of Sr, Y, and Zr relative to Ba are a few
orders of magnitude lower than what would be expected for stellar
models with $Z=0.5Z_\odot$. The $s$-process in low metallicity AGB
stars thus remains open to investigation, particularly with regard
to the magnitude of the $^{13}$C pocket. We are fully aware that
there have been remarkable advances in models of AGB evolution and
$s$-process nucleosynthesis over the past decade (see Busso et al.
1999 and references therein). Our purpose in the above discussion
is not to decry the status of that field but to call attention to
some uncertainties critical to our study.

The solar $r$-abundance of element E is calculated as
\begin{equation}
({\rm E/H})_{\odot,r}=({\rm E/H})_\odot - ({\rm E/H})_{\odot,s}
\equiv (1-\beta_{\odot,s})({\rm E/H})_\odot.
\end{equation}
In all cases where $\beta_{\odot,s}({\rm E})$ is small, the
results for (E/H)$_{\odot,r}$ are sound and provide the basis for
comparison of the observed $r$-abundance patterns in UMP and MP
stars with the solar $r$-component. In this work we are concerned
with Sr, Y, Zr, and Ba for which $\beta_{\odot,s}\sim 1$ with
substantial uncertainties (K\"appeler et al. 1989; Arlandini et
al. 1999). For example, the value for (Sr/H)$_{\odot,r}$ assigned
by K\"appeler et al. (1989) has an uncertainty of 111\% and is
larger by factors of $\approx 2$--3 than the values given by
Arlandini et al. (1999). For Y, the calculated
$s$-process contribution
exceeds its total solar inventory in the classical analysis
by Arlandini et al. (1999) and their stellar model calculations
give a value for (Y/H)$_{\odot,r}$ that is smaller by a factor of
3.5 than the value assigned by K\"appeler et al. (1989). Further,
the division of the solar inventory into $s$-components and
$r$-components is not well defined for Sr, Y, and Zr as these
elements can also be produced by the $\alpha$-process that does
not involve neutron capture but may occur in close connection with
the $r$-process in SNII (Woosley \& Hoffman 1992). In the case of
Ba, K\"appeler et al. (1989) gave only an upper limit for
($^{138}$Ba/H$)_{\odot,r}$ and Arlandini et al. (1999) have
already found shifts in the main $s$-component of Ba that result
in a range of $[1-\beta_{\odot,s}({\rm Ba})]$ from 0.08 in the
classical analysis to 0.19 in the stellar model calculations. This
clearly illustrates that care must be taken in using the solar
$r$-abundances when the corresponding $\beta_{\odot,s}$ values are
close to unity.

The $\beta_{\odot,s}$ values for the elements Y and above that are
used here correspond to the main $s$-component calculated by
Arlandini et al. (1999) from stellar models. These results
represent a significant step forward from the phenomenological
approach of the classical analysis but they do not represent a
full calculation of the integrated $s$-process
production by AGB stars
over Galactic history. In addition to the main $s$-component, the
classical analysis also introduces a so-called weak $s$-component
that is important for Sr and below (e.g., K\"appeler et al. 1989).
The $\beta_{\odot,s}$ value for Sr used here is also taken from
the stellar model calculations of Arlandini et al. (1999) but
includes the contribution from the weak $s$-component considered
by these authors. The $\beta_{\odot,s}$ values cited above and
adopted for our initial calculations will be shown to result in a
mismatch between the calculated Sr, Y, Zr, and Ba abundances and
the observational data on MP stars. It is widely recognized in the
field that the abundances of Sr, Y and Zr are problematic,
particularly at low metallicities, relative to any current
nucleosynthetic models (M. Busso \& R. Gallino, personal
communication). While we also explore other explanations for the
mismatch between our model results and data (see \S\ref{dis}), we
feel that it is reasonable to consider the uncertainties in the
solar $s$-component for Sr, Y, Zr, and Ba. It is possible that the
same uncertainties may also affect other elements whose 
$\beta_{\odot,s}$ values are not small.

\section{The $P$-Inventory and $H$ and $L$ Yields}
\label{yields}
The $P$-inventory and $H$ and $L$ yields are the key input to
our model. These are calculated in this section.

\subsection{The $P$ and $H$ Components}
\label{phcomp}
As addition of Fe from a single $L$ event to the
$P$-inventory would give [Fe/H]~$\approx -2.4$ (see eq.
[\ref{fe}]), UMP stars with [Fe/H]~$\approx -3$
could not have received any contributions
from $L$ events (i.e., $n_L=0$). Thus the abundances in these
stars reflect only the addition of $H$-contributions
to the $P$-inventory, providing a unique record for the
$P$ and $H$ components.
Equation (\ref{eps}) can be simplified for UMP stars as
\begin{equation}
\label{ump}
10^{\log\epsilon({\rm E})}=10^{\log\epsilon_P({\rm E})}
+n_H\times 10^{\log\epsilon_H({\rm E})}.
\end{equation}
The value $n_H$ for a star can be calculated from the observed Eu
abundance in the star by using equation (\ref{eu}). For example,
Eu data on the UMP stars HD 122563, HD 115444, and CS 22892-052
(Westin et al. 2000; Sneden et al. 2000) correspond to $n_H\approx 1$,
7, and 36 (see Table \ref{syzph}). The abundances
of element E other than Eu in two UMP stars with different $n_H$
provide a pair of equations in the form of equation (\ref{ump})
with two unknown parameters $\log\epsilon_P({\rm E})$ and
$\log\epsilon_H({\rm E})$. Thus the data on two such UMP stars
(singly and doubly primed quantities) can
be used to determine simultaneously the $P$-inventory and
$H$-yield of element E:
\begin{mathletters}
\begin{eqnarray}
10^{\log\epsilon_P({\rm E})}&=&{n_H'\times
10^{\log\epsilon''({\rm E})}-n_H''\times
10^{\log\epsilon'({\rm E})}\over n_H'-n_H''},\\
10^{\log\epsilon_H({\rm E})}&=&{10^{\log\epsilon'({\rm E})}-
10^{\log\epsilon''({\rm E})}\over n_H'-n_H''}.
\end{eqnarray}
\end{mathletters}
The results for Sr, Y, and
Zr obtained by using the data (see Table \ref{syzph}) on HD 115444
([Fe/H]~$=-2.99$) and CS 22892-052 ([Fe/H]~$=-3.1$) are given in
Table \ref{phl}.

The $P$-component of Sr, Y, and Zr is associated with the prompt
production of Fe. Thus any difference in [Fe/H] among UMP stars
should also be reflected by the $P$-component of these elements
(see below). This indicates that the $P$-inventory of Sr, Y, and
Zr for HD 122563 with [Fe/H]~$=-2.74$ should be $\approx 0.3$ dex
higher than the $\log\epsilon_P$ values for these elements
calculated above by using the data on HD 115444 and CS 22892-052,
which have lower [Fe/H] values (by $\approx 0.3$ dex). The
observed Eu abundance in HD 122563 gives $n_H\approx 1$ for this
star (see Table \ref{syzph}). The $H$-yields of Sr, Y, and Zr
calculated above are low, which suggests that the abundances of
these elements in HD 122563 should essentially reflect their
$P$-inventory for this star. The observed abundances of Sr, Y, and
Zr in HD 122563 (see Table \ref{syzph}) are indeed $\approx
0.2$--0.3 dex higher than the $\log\epsilon_P$ values in Table
\ref{phl}.

The level of metallicity necessary for the formation of the first
generation of normal stars is not known {\it a priori}. We have
inferred from the observational data that normal astration first
occurred at [Fe/H]~$\approx-3$ based on the phenomenological
model. Recently, Bromm et al. (2001) have shown that a gas cloud
with a baryon mass of $10^5\,M_\odot$ and $Z=10^{-4}Z_{\odot}$
fails to undergo continued collapse and fragmentation whereas for
$Z=10^{-3}Z_{\odot}$ the cloud can collapse and undergo vigorous
fragmentation, leading to a large relative population of low mass
clumps compared with the case of $Z=0$ (Big Bang debris). The
phenomenological and dynamical views thus appear to be in accord
for the moment. The difference in [Fe/H] among UMP stars referred
to in our above discussion may have resulted from a range of
values for the $P$-inventory of Fe, possibly as wide as $-4< {\rm
[Fe/H]}_P< -2.7$, over which normal astration may first have
occurred. We have chosen a nominal value of [Fe/H]$_P\approx -3$
in the above calculation of the $P$-component. As we assign no Eu
or any other heavy $r$-element to the $P$-inventory, the possible
range of [Fe/H]$_P$ does not affect the calculation of $n_H$ nor
that of $\log \epsilon({\rm E})$ for heavy $r$-elements produced
by $H$ events. However, elements such as Sr, Y, and Zr associated
with the $P$-inventory are directly related to the prompt Fe
production. It follows that $\log \epsilon_P({\rm E})$ for these
elements will be shifted with $\log \epsilon_P ({\rm Fe})$
depending on the precise value of [Fe/H] at which normal astration
began. In comparing the $P$-components of element E for UMP stars,
it is therefore necessary to correct to a standard value of [Fe/H]
in order to obtain the $P$-inventory of element E relative to Fe.
The $\log \epsilon_P({\rm E})$ values in Table \ref{phl} should be
associated with [Fe/H]$_P\approx -3$. If a UMP star has
[Fe/H]~$=-3+\delta$, then its $P$-component has to be shifted by
$\delta$ from the tabulated $\log \epsilon_P({\rm E})$ values as
was done above for HD 122563.

We have assigned the heavy $r$-elements above Ba (possibly
including Ba) exclusively to $H$ events with a yield template that
is identical to the corresponding part of the solar $r$-abundance
pattern (see \S\ref{heavy}). For these elements, including Eu
whose observed abundance is used to calculate $n_H$ for a star,
$\log \epsilon_P({\rm E})=-\infty$ and $\log\epsilon_H({\rm
E})=\log\epsilon_{\odot,r}({\rm E})- \log n_H^\odot$, where
$n_H^\odot=10^3$ is the number of $H$ events that contributed to
the solar inventory. The $H$-yields of Ba to Au calculated this
way have been given by QW and are repeated
in Table \ref{phl}. The $H$-yields of heavy $r$-elements other
than Eu may also be directly calculated from the observed
abundances of these elements in any UMP star without using their
solar $r$-inventory. For example, CS 22892-052 with $n_H^{\rm
CS}\approx 36$ (see Table \ref{syzph}) has $\log\epsilon_{\rm
CS}({\rm Ba})=-0.01$ (Sneden et al. 2000). This gives
$\log\epsilon_H({\rm Ba})= \log\epsilon_{\rm CS}({\rm Ba})- \log
n_H^{\rm CS}\approx -1.57$ in excellent agreement with the value
$\log\epsilon_H({\rm Ba})=\log\epsilon_{\odot,r}({\rm Ba})-
\log n_H^\odot=-1.52 $ obtained above for
$\log\epsilon_{\odot,r}({\rm Ba})=1.48$
(Arlandini et al. 1999). The long-lived chronometers $^{232}$Th and
$^{238}$U are produced along with the stable elements above Ba
(possibly including Ba) in $H$ events. The $H$-yields of these two
nuclei have been calculated by QW based
on their solar $r$-abundances and are given in Table \ref{phl}.
Of the stable elements above Au, only Pb has been observed in UMP
and MP stars with large uncertainties (e.g., Sneden et al. 2000).
The $H$-yields of these elements are not given here.

Besides Sr, Y, and Zr, the light $r$-elements Nb, Ru, Rh, Pd, Ag,
and Cd have been observed in the UMP star CS 22892-052, but with
significantly lower abundances relative to the solar $r$-abundance
pattern that is translated to pass through the Eu data (Sneden et
al. 2000). This suggests that some light $r$-elements may be
produced along with the heavy $r$-elements in $H$ events, but with
relative yields of light to heavy $r$-elements significantly
smaller than the corresponding solar $r$-abundance ratios. If we
consider that the $P$-inventory of Nb, Ru, Rh, Pd, Ag, and Cd
is overwhelmed by contributions from the $n_H^{\rm CS}\approx 36$
$H$ events assigned to CS 22892-052 (see Table \ref{syzph}),
then the $H$-yields of these
elements can be estimated from their observed abundances,
$\log\epsilon_{\rm CS}({\rm E})$, as:
$\log\epsilon_H({\rm E})\approx\log\epsilon_{\rm CS}({\rm E})-\log
n_H^{\rm CS}$. Similar results were obtained by QW
who used the observed abundance of
Os instead of Eu to calculate $n_H^{\rm CS}$.
The $P$-inventory and
$H$-yields derived above are summarized in Table \ref{phl} and
shown in Figure \ref{figphl}.

\subsection{The $L$ Component}
\label{lcomp} 
As discussed in \S\ref{phcomp}, the $P$-inventory
and $H$-yields are calculated from data on two UMP stars with
different $n_H$ values. In this calculation, only the solar
$r$-inventory of Eu is needed to obtain the $H$-yield of Eu. The
yields for the other elements can be directly obtained from
observations of UMP stars although the yields of the other heavy
$r$-elements can also be calculated
by using their solar $r$-inventory.
In contrast, the $L$-yields will be
calculated by using the solar $r$-inventory as an essential input.
This inventory was contributed by $n_H^\odot=10^3$ $H$
and $n_L^\odot=10^2$ $L$ events in addition to the $P$-inventory:
\begin{equation}
\label{sun} ({\rm E/H})_{\odot,r}=({\rm E/H})_P+n_H^\odot({\rm
E/H})_H+ n_L^\odot({\rm E/H})_L.
\end{equation}
The solar $r$-abundance $\log\epsilon_{\odot,r}({\rm E})$ given in
Table \ref{phl} (Arlandini et al. 1999; see \S\ref{rsun}) is used
in equation (\ref{sun}) to calculate the $L$-yield of element E
from the $\log\epsilon_P({\rm E})$ and $\log\epsilon_H({\rm E})$
values in Table \ref{phl}. The resulting $L$-yields from this
calculation are also given in Table \ref{phl} and shown in Figure
\ref{figphl}. Note that $\log\epsilon_L({\rm E})=-\infty$ for the
heavy $r$-elements above Ba (possibly including Ba)
as they have been exclusively assigned
to $H$ events (see \S\ref{phcomp}).

\section{Comparison of Model Results with Observational Data}
\label{compare}
The $\log\epsilon_P({\rm E})$, $\log\epsilon_H({\rm E})$, and
$\log\epsilon_L({\rm E})$ values in Table \ref{phl}
permit us to calculate the abundance
of element E in any UMP or MP star from the model described in
\S\ref{intro}
by using only the observed Eu and Fe abundances to obtain $n_H$ and
$n_L$ for the star (see eqs. [\ref{eps}], [\ref{eu}], and [\ref{fe}]).
The calculated abundances are compared with the observational data
in this section.

\subsection{Abundances of Ba and Above in UMP and MP Stars}
\label{heavy}
The comparison of model results with observational
data is the simplest for the heavy $r$-elements above Ba (possibly
including Ba) as they have been exclusively assigned to $H$ events
with a yield template identical to the corresponding part of the
solar $r$-abundance pattern (see \S\ref{phcomp}). As a test of
this hypothesis, we consider the abundance ratios of other heavy
$r$-elements (E) to Eu. These ratios should remain at the
solar $r$-process values independent of [Fe/H] (i.e., the
occurrence of $L$ events) if the heavy $r$-elements come only
from $H$ events. Based on the data of Burris et al. (2000; open
squares), Johnson \& Bolte (2001; asterisks), and Gratton \& Sneden
(1994; filled triangles), the difference
between the observed E/Eu and the solar $r$-process value of
Arlandini et al. (1999),
[E/Eu]$_r\equiv\log({\rm E/Eu})-\log({\rm E/Eu})_{\odot,r}$, is
shown as a function of [Fe/H] for Ba, La, Ce, Pr, Nd, Sm, Gd, and Dy in
Figure \ref{heavyeu}. The data sets of Burris et al. (2000) and
Johnson \& Bolte (2001) have 17 stars in common. Figure
\ref{heavyeu} shows that: (1) Ba/Eu exhibits a regular increase with
[Fe/H]; (2) La/Eu appears to lie within $\approx 0.3$ dex of the solar
$r$-process value over $-3\lesssim[{\rm Fe/H}]\lesssim -1$
(with the exception of several data points); 
(3) Nd/Eu shows a wide scatter
with no identifiable trend; (4) Dy/Eu appears to lie within
$\approx 0.3$ dex of the solar $r$-process value over
$-3\lesssim[{\rm Fe/H}]\lesssim -1$
(with the exception of several data points); (5) Ce/Eu, Pr/Eu,
Sm/Eu, and Gd/Eu have only sparse data with no indication of
any wide spread from the solar $r$-process values. 
In general, all those elements above Ba
(except for Nd) appear to have ratios relative to Eu
that are roughly constant. This demonstrates that they are to be
associated exclusively with $H$ events,
which is in accord with the model. We cannot explain the scatter
in Nd/Eu. Further, we note that the observed
values for La/Eu, Ce/Eu, Pr/Eu,
Sm/Eu, Gd/Eu, and Dy/Eu are close to the standard solar
$r$-process values. Of these elements, Sm, Eu, Gd, and Dy have
predominant $r$-process contributions to their solar inventory
$(\beta_{\odot,s}\ll 1$; see Table \ref{phl}). However, La has
$\beta_{\odot,s}({\rm La}) = 0.62$ from stellar model calculations
and $\beta_{\odot,s}({\rm La}) = 0.83$ from the classical analysis
(Arlandini et al. 1999). These $\beta_{\odot,s}({\rm La})$
values correspond to a change in the solar
$r$-component of La by a factor of 2.2. Thus the present
agreement between the observed La/Eu and the corresponding solar
$r$-process value from stellar model calculations of 
Arlandini et al. (1999)
is surprising, but encouraging. As La, Ce, and Pr have
$\beta_{\odot,s}\gtrsim 0.5$ (see Table \ref{phl}), 
it is clear that any
significant $s$-process addition to the ISM would drastically
shift their abundances away from the nominal solar $r$-process values.
This is not observed. We conclude that the observational data appear to
justify two basic assumptions of our model: the $H$ events dominate
the production of heavy $r$-elements from La and above and there is no
significant $s$-process contribution to the ISM prior to
[Fe/H]~$\sim -1$. The new data also support and extend the earlier
observational results that the $r$-abundance pattern above Ba in
UMP stars is the same as that in the solar system (Sneden et al.
1996, 2000; Westin et al. 2000). The need for more precise observational
data on La and above is evident.

Figure \ref{heavyeu} shows that Ba/Eu exhibits
systematic deviation from a constant value
with increasing [Fe/H]. Johnson \& Bolte (2001) have
emphasized that the Ba abundance data may reflect uncertainties
in stellar atmosphere models. These uncertainties may have caused
the deviation for Ba/Eu as well as the wide scatter in
Nd/Eu discussed above.
However, the extent to which artifacts in stellar atmosphere
models would affect the abundance analysis for Ba and other elements
remains to be investigated. On the other hand,
Ba is special in that it is closer to the boundary between
light and heavy $r$-elements. This boundary has been considered to
be at $A\sim 130$ (WBG). It was also noted that any $r$-process
model is expected to produce the abundance peak at $A\sim 130$
with some overflow to higher masses (Qian et al. 1998). In this
case, some contributions may be expected at and above Ba from the
$L$ events that are mainly responsible for the light $r$-elements.
A test of such overflow would be shifts in the abundance ratios of
other heavy $r$-elements to Eu with increasing $L$-contributions
(i.e., increasing [Fe/H]). As discussed above, no such shifts have
been observed for La (see Figure \ref{heavyeu}). Thus the overflow
in $L$ events must stop below La ($A=139$). But Ba is below La and
may be liable to the influence of such overflow. So the
deviation of Ba/Eu from a constant value with increasing [Fe/H]
may be caused by the $L$-contributions to Ba.
This possibility will be discussed further in \S\ref{syzbmp}.

\subsection{Abundances of Sr, Y, Zr, and Ba in UMP stars}
\label{syzbump}
The $P$-inventory and $H$-yields of Sr, Y, and Zr in Table
\ref{phl} are calculated from data on HD 115444 and CS 22892-052
(both with [Fe/H]~$\approx -3$). These results are used to
calculate the Sr, Y, and Zr abundances in four other UMP stars in
the sample of Burris et al. (2000). For example, BD
+58$^\circ$1218 has $\log\epsilon({\rm Eu})=-2.00$ and
[Fe/H]~$=-2.72$ corresponding to $n_H\approx 3$ and $n_L\approx
0$. As its [Fe/H] differs from [Fe/H]$_P\approx -3$ by
$\delta\approx 0.28$ dex, the corresponding $P$-components of Sr,
Y, and Zr must be increased from the $\log\epsilon_P$ values in
Table \ref{phl} by the same amount (see \S\ref{phcomp}). The
calculated Sr abundance for this star is $\log\epsilon_{\rm
cal}({\rm Sr})= \log[10^{\log\epsilon_P({\rm
Sr})+\delta}+n_H\times 10^{\log\epsilon_H({\rm Sr})}]\approx
0.43$. The values of $\log\epsilon_{\rm cal}({\rm Y})\approx
-0.71$ and $\log\epsilon_{\rm cal}({\rm Zr})\approx 0.18$ are
obtained similarly. These calculated abundances are within
$\approx 0.2$ dex of the observational data $\log\epsilon_{\rm
obs}({\rm Sr})=0.35$, $\log\epsilon_{\rm obs}({\rm Y})=-0.58$, and
$\log\epsilon_{\rm obs}({\rm Zr})=-0.06$. The difference between
the model results and data, $\Delta\log\epsilon({\rm E})\equiv
\log\epsilon_{\rm cal}({\rm E})-\log\epsilon_{\rm obs}({\rm E})$,
is shown for the four UMP stars in the sample of Burris et al.
(2000) in Figure \ref{figsyzb} (filled squares). Except for Sr in
HD 126587, $|\Delta\log\epsilon({\rm E})|\lesssim 0.2$ dex for
all other cases. We note that McWilliam et al. (1995) gave
$\log\epsilon_{\rm obs}({\rm Sr})=0.22$ for HD 126587, which is
0.62 dex higher than the value of Burris et al. (2000). Using the
value of McWilliam et al. (1995) would bring the calculated Sr
abundance of this star within 0.2 dex of the data.

With regard to Ba in UMP stars, we have calculated the abundances
assuming pure $H$ contributions and using the observed
$\log\epsilon({\rm Eu})$ to determine the value of $n_H$. The
calculated values are in very good agreement with the
observational data (see filled squares in Figure \ref{figsyzb}). 
There are two issues regarding the Ba data at
$-4\lesssim$~[Fe/H]~$<-2.5$ (see Figure \ref{ba}). The low
Ba abundances at $-4\lesssim$~[Fe/H]~$<-3$ compared with the
$H$-yield of Ba (dotted line in Figure \ref{ba}) were attributed
to very small admixtures of $H$ contributions into clouds having
almost pure $P$-inventory composition. There is also a rapid rise
of $\log\epsilon({\rm Ba})$ at $-3\lesssim$~[Fe/H]~$<-2.5$. The
dot-dashed line from the model (see \S\ref{syzbmp}) that describes
the trend of data at [Fe/H]~$>-2.5$ where both $H$ and $L$ events
occur is extended to [Fe/H]~$=-4$ in Figure \ref{ba}. The
deviation (by $\sim 1$--2 dex) of the data at
$-3\lesssim$~[Fe/H]~$<-2.5$ from the tend at [Fe/H]~$>-2.5$ is
evident. This is the wall of rapid rise described by WQ and was
attributed to the onset of $H$ events with no Fe production. A
problem exists as to how high this rise might be if it represents
a sampling of the ISM. As the $H$ yield of Ba,
$\log\epsilon_H({\rm Ba})\approx -1.5$, is well established and
the frequency of $H$ events is $\sim (10^7\ {\rm yr})^{-1}$, it is
clear that for a UMP star, $\log\epsilon({\rm Ba})$ should not
represent more $H$ events than say $n_H\sim 20$ corresponding to a
period of $\sim 2\times 10^8$~yr during which an $L$ event should
occur with very high probability. Observed values of $\log\epsilon
({\rm Ba})$ representing $n_H\gtrsim 30$ $H$ events but with no
concomitant increase in [Fe/H], such as in the cases of CS
31082-001 (crosses linked by a line in Figure \ref{ba}) and CS
22892-052 (open circle), are then attributed to contamination by
an SNII $H$ event in a binary and not to gross increases  of
$r$-abundances in the ISM from which the stars formed (QW). 
Surface contamination may also
apply to some data at [Fe/H]~$>-2.5$ that lie far above the trend
line in Figure \ref{ba}. Clearly, chemical enrichments
by surface contamination in binary
systems cannot be ignored (see Preston \& Sneden 2001).

\subsection{Abundances of Sr, Y, Zr, and Ba in MP stars}
\label{syzbmp} 
Here we focus on the discussion of MP stars. The
range of $-3<[{\rm Fe/H}]<-1$ for MP stars is considered to result
from addition of Fe from $L$ events to the $P$-inventory. In
extending the discussion from UMP stars with no contributions from
$L$ events to MP stars with such contributions, estimates of the
$L$-yields play an essential role. The $L$-yields,
$\log\epsilon_L({\rm E})$, are calculated from the values of
$\log\epsilon_{\odot,r}({\rm E})$, $\log\epsilon_P({\rm E})$, and
$\log\epsilon_H({\rm E})$ given in Table \ref{phl}. The abundances
of Sr, Y, and Zr are then calculated from the model by using the
$P$-inventory and $H$ and $L$ yields together with the observed
$\log\epsilon({\rm Eu})$ and [Fe/H] for the MP stars in the sample
of Burris et al. (2000). For example, HD 122956 has
$\log\epsilon({\rm Eu})=-0.72$ and [Fe/H]~$=-1.78$ corresponding
to $n_H\approx 58$ and $n_L\approx 5$. The calculated Sr abundance
for this star is $\log\epsilon_{\rm cal}({\rm Sr})=
\log[10^{\log\epsilon_P({\rm Sr})}+n_H\times
10^{\log\epsilon_H({\rm Sr})}+n_L\times 10^{\log\epsilon_L({\rm
Sr})}]\approx 0.63$. The values of $\log\epsilon_{\rm cal}({\rm
Y})\approx -0.08$ and $\log\epsilon_{\rm cal}({\rm Zr})\approx
0.65$ are obtained similarly. By comparison, the observed values
are $\log\epsilon_{\rm obs}({\rm Sr})=1.30$, $\log\epsilon_{\rm
obs}({\rm Y})=0.28$, and $\log\epsilon_{\rm obs}({\rm Zr})=0.98$.
Thus the calculated abundances (especially for Sr) significantly
differ from the data. The difference between the model results and
data, $\Delta\log\epsilon({\rm E})\equiv \log\epsilon_{\rm
cal}({\rm E})-\log\epsilon_{\rm obs}({\rm E})$, is shown as a
function of [Fe/H] for the MP stars in the sample of Burris et al.
(2000) in Figure \ref{figsyzb} (open squares). It can be seen that
the model systematically underestimates Sr, Y, and Zr abundances
at $-2.5\lesssim{\rm [Fe/H]}< -1$. As these metallicities
correspond to increasing contributions from $L$ events, we
consider that this systematic deviation is possibly caused by
errors in the $L$-yields of Sr, Y, and Zr as determined from the
solar $r$-inventory.

Note that the solar $r$-abundance of Sr assigned by Arlandini et
al. (1999) is saturated by the contributions from
$n_H^\odot=10^3$ $H$ events and thus
does not allow any $L$-contributions to Sr. However, the 
solar $r$-abundances of Y and Zr assigned
by the same authors allow substantial
$L$-contributions to these two elements (see Table \ref{phl}). As
Sr, Y, and Zr are close in atomic mass, it is rather difficult to
explain why $L$ events can produce Y and Zr but no Sr. It seems
more likely that the standard solar $r$-abundances of Sr, Y, and
Zr are too uncertain to give reliable $L$-yields for these
elements. As discussed in \S\ref{rsun}, the solar $r$-process
fraction $[1-\beta_{\odot,s}({\rm E})]$ for element E can have
quite large errors when $\beta_{\odot,s}({\rm E})$ is close to
unity as for Sr, Y, and Zr (see Table \ref{phl}). We now consider
that the solar $r$-inventory of Sr, Y, and Zr given by Arlandini
et al. (1999) and adopted above may be increased by factors of
6.3, 3.5, and 3, respectively, to obtain a ``corrected'' solar
$r$-inventory $\log\epsilon_{\odot,r}^{\rm corr}({\rm E})$ and the
corresponding $\log\epsilon_L^{\rm corr}({\rm E})$ values in Table
\ref{phl}. With these $\log\epsilon_L^{\rm corr}({\rm E})$ values,
the model results on Sr, Y, and Zr abundances at $-2.5\lesssim{\rm
[Fe/H]}< -1$ are generally in quite good agreement with data (see
filled circles in Figure \ref{figsyzb}). The problems and issues
that are raised by the hypothesized shifts in
$\beta_{\odot,s}({\rm E})$ for Sr, Y, and Zr will be discussed in
\S\ref{dis}. It has been well known for some
time that the abundances of Sr, Y, and Zr in UMP and MP stars 
cannot be explained by any standard nucleosynthetic models 
(M. Busso
\& R. Gallino, personal communication). We note that the
$P$-inventory, $H$ yields, and corrected $L$ yields of these
elements given here follow approximately the same relative
production pattern with a significant dip at Y. This is not the
case if the standard $\beta_{\odot,s}$ values for Sr, Y, and Zr
are used.

If we assign $r$-process production of Ba exclusively to $H$
events, the Ba abundance in an MP star can be calculated as
$\log\epsilon_{\rm cal}({\rm Ba}) =\log\epsilon_H({\rm Ba})+ \log
n_H$, where $\log\epsilon_H({\rm Ba})$ can be obtained from the
solar $r$-inventory of Ba for $n_H^\odot=10^3$ or from the data on
CS 22892-052 with essentially the same result (see \S\ref{phcomp}
and Table \ref{phl}). The difference between the calculated
results and data of Burris et al. (2000), $\Delta\log\epsilon({\rm
Ba}) \equiv \log\epsilon_{\rm cal}({\rm Ba})-\log\epsilon_{\rm
obs}({\rm Ba})$, is shown as a function of [Fe/H] in Figure
\ref{figsyzb} (open squares). Note that $\Delta\log\epsilon({\rm
Ba})$ lies significantly below zero at $-2.5\lesssim{\rm [Fe/H]}<
-1$. This is the same as found for Sr, Y, and Zr. As indicated
above, this result follows whether we calculate
$\log\epsilon_H({\rm Ba})$ by using the data on a UMP star or by
attributing the standard solar $r$-inventory of Ba 
from Arlandini et
al. (1999) to $n_H^\odot=10^3$ $H$ events. If we consider that the
solar $r$-inventory of Ba assumed above is too small and that $L$
events contribute some Ba, we may find a change that could
eliminate the discrepancy. Using $\log\epsilon_H({\rm Ba})\approx
-1.57$ calculated from the data on CS 22892-052 (see
\S\ref{phcomp}) and increasing the solar $r$-inventory of Ba by a
factor of 2 from the value given by Arlandini et al. (1999), we
obtain $\log\epsilon_L^{\rm corr}({\rm Ba})\approx -0.47$ (see eq.
[\ref{sun}]). We then recalculate $\log\epsilon_{\rm cal}({\rm
Ba})$ from equation (\ref{eps}) by including the contributions to
Ba from $L$ events. The recalculated $\Delta\log\epsilon({\rm
Ba})$ is shown in Figure \ref{figsyzb} (filled circles). It can be
seen that the model results with Ba contributions from $L$ events
represent data quite well over $-2.5\lesssim{\rm [Fe/H]}< -1$. This
also accounts for the trend of Ba/Eu (see \S\ref{heavy} and Figure
\ref{heavyeu}) found by Burris et al. (2000) and Johnson \& Bolte
(2001): the shift in Ba/Eu with [Fe/H] is caused by the
increasing contributions to Ba from $L$ events
that do not produce Eu. The mean trend for
evolution of Ba abundance relative to Fe at [Fe/H]~$>-2.5$ can be
calculated from the values of $\log\epsilon_H({\rm Ba})$,
$\log\epsilon_L^{\rm corr}({\rm Ba})$, and  $\log\epsilon_L({\rm
Fe})$ by assuming $n_H/n_L=10$. This is shown as the dot-dashed
line in Figure \ref{ba}.

As an independent test, we may calculate $\log\epsilon_L({\rm Ba})$
from data on HD 126238 (Gratton \& Sneden 1994) that is not in
the sample of Burris et al. (2000). In this calculation the part
of the solar Fe inventory contributed by SNII is used to determine
the $L$-yield of Fe, and hence $n_L$, but the solar
$r$-inventory of Ba is not used. The value of $\log\epsilon_L
({\rm Ba})$ is calculated from the value of
$\log\epsilon_H({\rm Ba})$ (obtained from the data on CS 22892-052)
and the observed value of $\log\epsilon({\rm Ba})$ in HD 126238
together with the values of $n_H$ and $n_L$ for this star.
The observed values of
$\log\epsilon({\rm Eu})=-0.87$ and [Fe/H]~$=-1.67$ for HD 126238
correspond to $n_H\approx 41$ and $n_L\approx 6$. Using
$\log\epsilon_H({\rm Ba})\approx -1.57$, we obtain
$\log\epsilon_L({\rm Ba})\approx -0.46$ from the observed value of
$\log\epsilon({\rm Ba})=0.50$ for HD 126238.
This is in excellent agreement with
the above value of $\log\epsilon_L^{\rm corr}({\rm Ba})
\approx -0.47$. It thus
appears that the value of $\log\epsilon_L({\rm Ba})$ directly
determined from data on a single MP star not in the sample of
Burris et al. (2000) without using the solar $r$-inventory of Ba
is in accord with the value
obtained by changing this inventory to fit the data of 
Burris et al. (2000)
on a large sample of MP stars. We consider the above direct calculation
and the problem of fitting the Ba data as some evidence indicating
that the standard solar $r$-inventory of Ba may require revision.
The required change in this inventory by a factor of 2 is not within
the $1\sigma$ error band for $[1-\beta_{\odot,s}({\rm Ba})]$ given by
Arlandini et al. (1999), but corresponds to a change of only 23\%
from their assigned value of $\beta_{\odot,s}({\rm Ba})=0.81$ to
the corrected value of $\beta_{\odot,s}^{\rm corr}({\rm Ba})=0.62$.

To further test the model with corrected solar $r$-inventory of
Sr, Y, Zr, and Ba, we apply it to the data sets of
Magain (1989) and Gratton \& Sneden (1994) that are distinct from
the data set of Burris et al. (2000). The difference between the
model results and data, $\Delta\log\epsilon({\rm E}) \equiv
\log\epsilon_{\rm cal}({\rm E})-\log\epsilon_{\rm obs}({\rm E})$,
is shown for Sr, Y, Zr, and Ba as a function of [Fe/H] in Figure
\ref{test}. It can be seen that by using the corrections made to 
the solar $r$-inventory for these elements based on
the data set of Burris et al. (2000), the model  
describes the data sets of
Magain (1989) and Gratton \& Sneden (1994) equally well
($|\Delta\log\epsilon({\rm E})|\lesssim 0.2$~dex in most cases).
Thus we consider that the
model presented above at least represents a very good parametric
fit to the observational data on UMP and MP stars from several
groups. It also seems possible that the model has deeper
implications for $s$-process and $r$-process nucleosynthesis
models and for the chemical evolution of the Galaxy.

\section{Discussion and Conclusions}
\label{dis}
We have developed the consequences of a phenomenological model for
the abundances of elements in the ISM for [Fe/H]~$< -1$. This model
assumes three components of nucleosynthetic contributions to
$r$-process related elements. One
component is contributed by the first generations of very
massive ($\gtrsim 100\,M_{\odot}$) stars formed after the Big Bang.
These stars provide an initial or prompt inventory of Fe and
associated elements corresponding to [Fe/H]$_P\approx -3$. This
$P$-inventory is distinctive and contains no heavy $r$-elements
(with $A>130$). Once the $P$-inventory is established,
normal stellar populations with masses of $\sim (1$--$60)\,M_\odot$,
which include the progenitors for
two types of SNII, begin to form at [Fe/H]~$\approx -3$. The two
types of SNII correspond to the other two components of the model.
They are the high frequency $H$ events that produce mainly heavy
$r$-elements but no Fe, and the low frequency $L$ events that
produce mainly light $r$-elements (with $A\lesssim 130$) and Fe.
The composition of the $P$-inventory and the yield templates of
$H$ and $L$ events are assumed to be fixed and distinct from each
other. This
allows the deconvolution of observational data to establish
quantitatively the $P$-inventory and $H$ and $L$ yields. The
basic input for this procedure is the observational data on two
UMP stars and the solar $r$-abundances.

The element Eu commonly observed in UMP and MP stars is essentially
a pure $r$-element. As Eu and all the other heavy $r$-elements
above Ba are attributed exclusively to $H$ events, it is convenient
to use the observed Eu abundance in a UMP or MP star to identify
the $H$ contributions to the abundances in the star (any other 
heavy $r$-element above Ba can also be used for this purpose). 
The $H$-yield
of Eu is calculated from its solar $r$-abundance by using the
number of contributing $H$ events for the solar inventory, which
can be estimated from the frequency of these events. The number
of contributing $H$ events for a UMP or MP star can then be obtained
from the observed Eu abundance in the star by using the $H$-yield
of Eu. For UMP stars with [Fe/H]~$\approx$~[Fe/H]$_P\approx -3$,
their abundances only consist of $P$ and $H$ contributions.
By using the data on two UMP stars with [Fe/H]~$\approx -3$ but
with different Eu abundances, the $P$-inventory and $H$-yields of
Sr, Y, and Zr are simultaneously obtained. The data on a third
UMP star give essentially the same results for the $P$-inventory 
when corrected to the same [Fe/H]$_P$ value (see \S\ref{phcomp}). 
These results are used
in the model to calculate the Sr, Y, and Zr abundances in other
UMP stars. The calculated abundances are shown to be in very good
agreement with the observational data (see filled squares in
Figure \ref{figsyzb}). The $P$-inventory represents the sum of
contributions from the first generations of very massive stars.
We consider that the $P$-inventory of Sr, Y, and Zr reported here
is reliable and may serve as a guide in evaluating models of
nucleosynthesis for these stars.

The $H$-yields of heavy $r$-elements above Ba other than Eu can be
calculated from the solar $r$-abundances as in the case of Eu or
from the observed abundances in a UMP star by using the observed
Eu abundance to obtain the number of contributing $H$ events for
the star. These two approaches give effectively the same results
as the $r$-abundance pattern above Ba observed in UMP stars is the
same as that
in the solar system. For MP stars with $-3 < [{\rm Fe/H}]
< -1$, their Fe abundances reflect contributions from $L$ events.
In general, the abundances in these stars are governed by $P$,
$H$, and $L$ contributions. For the heavy $r$-elements above Ba
that only have $H$ contributions in our model, the abundance
ratios of the other heavy $r$-elements to Eu in MP stars are close
to the solar $r$-process values assigned by Arlandini et al.
(1999) over a wide range in [Fe/H] (see Figure \ref{heavyeu}). For
other elements that have $L$ contributions, their $L$-yields are
estimated from their solar $r$-abundances, $P$-inventory, and
$H$-yields by using the numbers of contributing $H$ and $L$ events
for the solar inventory. The abundances of these elements in an MP
star can then be calculated from the three-component model by
using the observed Eu and Fe abundances to obtain the numbers of
contributing $H$ and $L$ events for the star. By comparing the
model results with the observational data on $\sim 30$--40 MP
stars in the sample of Burris et al. (2000), it is found that to
match the Sr, Y, Zr, and Ba data requires major changes in the
solar $r$-abundances of these elements assigned by Arlandini et
al. (1999) (see Table \ref{phl} and Figure \ref{figsyzb}). The
model with a set of corrected solar $r$-abundances for these
elements determined by fitting the data set of Burris et al.
(2000) is shown to describe the two different data sets of Magain
(1989) and Gratton \& Sneden (1994) equally well over a wide range
in [Fe/H] (see Figure \ref{test}). All of the data sets on Ba
exhibit increases of Ba/Eu above the standard solar $r$-process
value with increasing [Fe/H]. This can be explained by the 
increasing contributions to Ba from $L$ events that do not produce 
Eu based on the model with the corrected solar $r$-abundance of Ba.

In summary, by using the $P$-inventory, $H$ yields, and $L$ yields
(see Table \ref{phl} and Figure \ref{phl})
that are obtained from the data on two UMP stars, the corrected
solar $r$-abundances for Sr, Y, Zr, and Ba, and the standard solar
$r$-abundances of Arlandini et al. (1999) for other elements, it is
possible to calculate the abundances of a large number of elements
in a UMP or MP star simply from the observed Eu and Fe abundances
in the star over the range $-3\lesssim {\rm [Fe/H]}<-1$.

The fundamental question arises as to whether the corrected solar
$r$-abundances for Sr, Y, Zr, and Ba can be justified in terms of
nuclear astrophysical considerations. All the other elements do
not seem to require a significant shift in their solar
$r$-inventory. It is shown that the model is robust insofar as the
relative yield patterns for $H$ and $L$ events are fixed. Changes
in the absolute yields 
(but not the relative yields) do not change the
patterns but only the numbers of contributing $H$ and $L$ events
for a star. The requirement for a large shift in the solar
$r$-inventory for Sr, Y, Zr, and Ba then remains unchanged in this
case. With regard to contributions from other sources, we note
that the observational data on La, Ce, and Pr with solar 
$s$-process fractions of $\beta_{\odot,s}\gtrsim 0.5$
show no evidence of
significant $s$-process contributions to the ISM at [Fe/H]~$< -1$.
Thus possible $s$-process additions to the ISM from AGB stars are
unimportant at [Fe/H]~$< -1$ and cannot affect the required
changes in the solar $r$-inventory for Sr, Y, Zr, and Ba. We note
that extremely high $s$-process enrichment has been observed in LP
625-44 with [Fe/H]~$=-2.7$ (Aoki et al. 2000). However, this
$s$-process enrichment does not reflect the composition of the ISM
at [Fe/H]~$=-2.7$ from which the star formed but is due to surface
contamination by mass transfer from the binary companion of the
star. Further, the observed [Fe/H] in the star only indicates the
formation time of the binary but does not correspond to the much
later occurrence of $s$-processing in the AGB companion.

It is possible that some other aspects of the model not considered
in the preceding discussion cause the disagreement between the
model results and observational data for Sr, Y, Zr, and Ba when
the ``standard'' solar $r$-inventory for these elements is used.
For example, the relative yields of the pertinent elements may
not be fixed as assumed but may vary from one $H$ or $L$ event
to another. In this case, the $H$ and $L$ contributions to the
abundances in a star cannot be identified simply by using the
observed abundances of Eu and Fe in the star. The problem of 
abundances in
UMP and MP stars would then require a totally different approach.
Unless confronted with new data that indicate otherwise, 
we adhere to the basic assumption
of constant relative yields for $H$ or $L$ events and consider
the following obvious alternatives to explain the disagreement 
referred to above: (1)
there are some systematic shifts in the observational data
provided by several groups for Sr, Y, Zr, and Ba as a function of
[Fe/H], perhaps due to uncertainties in the stellar atmosphere
models as suggested by Johnson \& Bolte (2001) for Ba; or (2)
there are intrinsic errors in the solar $r$-inventory for Sr, Y,
Zr, and Ba. 

The observed Ba/Eu shows a regular increase above
the standard solar $r$-process value with increasing
[Fe/H]. However,
the $H$-yield of Ba calculated from the data on a UMP
star is in exact agreement with its solar $r$-abundance assigned
by Arlandini et al. (1999) if there are no $L$ contributions to
Ba. This suggests that the observed shifts in Ba/Eu with
increasing [Fe/H] are the result of an artifact in abundance
analysis that also depends on [Fe/H]. We find it difficult to
understand how artifacts in analysis of the
observational data would cause large shifts in abundance only 
for Ba
without having large effects on other elements. If the Ba problem
is due to such an artifact, then the remaining inconsistency for
Sr, Y, and Zr may be accounted for by the acceptable $s$-process
uncertainties, and hence errors in the solar $r$-inventory for
these elements. We emphasize that for Sr, Y, and Zr, the
$\beta_{\odot,s}$ values calculated from $s$-process models are
all close to unity and lie in a region where serious uncertainties
or errors may exist as has long been recognized by experts in
$s$-process modeling. If the Ba problem is not due to  artifacts
in the abundance analysis, then the matter is not simply clarified
without changing the standard solar $r$-abundance of Ba. 
The case of Ba is
similar to that of Sr, Y, and Zr as $\beta_{\odot,s}({\rm Ba})$ is
also close to unity. However, Ba lies in a region where the main
$s$-component is dominant and is subject to more constraints than
Sr, Y, and Zr. At present, the solution to this problem is not
evident.  However, there is a legitimate basis to  consider that
the solar $r$-component obtained by subtracting the $s$-component
is subject to real uncertainties for elements such as Sr, Y, Zr,
and Ba with predominant $s$-process contributions to their solar
inventory. The solar $s$-component used here corresponds to
an average
of the $s$-process yields from models of $1.5\,M_\odot$
and $3\,M_\odot$ AGB stars with $Z=0.5 Z_{\odot}$ that best
reproduces the classical main $s$-component 
(Arlandini et al. 1999). In fact, the solar $s$-component is the 
integral of long term Galactic chemical
evolution starting with low metallicity AGB
stars. Researchers in the field are fully aware of this matter.
The solar $r$-component derived from the subtraction procedure
inherits all the uncertainties in the $s$-component calculated
from models. A reliable {\it ab initio} calculation of
the solar $r$-component based on models of SNII and Galactic
chemical evolution remains to be pursued. Critical attention to 
the investigation of $s$-process and $r$-process nucleosynthesis
indicated above is required to sort out the problems.

In any case, based on the observations of UMP stars, the
$P$-inventory produced by the first generations of very massive
stars and yields of SNII $H$ events appear to be well established.
This constitutes a prediction of what nucleosynthetic models for
very massive zero-metallicity stars and for the common SNII should
produce. By using the three-component model with corrected solar
$r$-abundances for Sr, Y, Zr, and Ba, and the standard values for
other elements, the abundances of a large number of elements in
stars with $-3\lesssim$~[Fe/H]~$< -1$ 
can be calculated very simply from the
observed Eu and Fe abundances. Whether this model is a reasonable
physical description of Galactic chemical evolution for
$r$-elements or is a parametric description with alternative
explanations cannot be answered at present.

\acknowledgments
We wish to thank the reviewers Achim Weiss and an
unidentified person for their comments and for suggesting a more
extensive treatment of the model. In particular, Achim Weiss has
raised several issues requiring discussion. A new
data set that appeared after the original submission and two
additional data sets are now treated.
We have profited greatly from comments and continued enthusiastic
support by Roger Blandford and the interest of Marc Kamionkowski.
Maurizio Busso has generously given aid and advice on the
$s$-process problems considered here. A conversation with Wal
Sargent stimulated our thinking about the mixing problem. This
work was supported in part by DOE grants DE-FG02-87ER40328 and
DE-FG02-00ER41149 (Y.Z.Q.) and by NASA grant NAG5-4083 (G.J.W.),
Caltech Division Contribution 8741(1072).

\clearpage
\figcaption{Data on $\log\epsilon({\rm Ba})$ 
(filled diamonds: McWilliam et al. 1995; McWilliam 1998;
open triangles: Westin et al. 2000; open squares: Burris et al. 2000;
open circle: Sneden et al. 2000) shown as a function of [Fe/H].
The upper cross indicates the Ba
abundance inferred for CS 31082-001 by QW based on the three-component
model and is within $\approx 0.3$ dex of the preliminary Ba data on
this star (lower cross) reported by Hill et al. (2001). The dotted
line labeled ``1 $H$ event'' indicates the Ba yield of a single $H$
event. The low Ba abundances at $-4\lesssim {\rm [Fe/H]}<-3$ 
compared with this $H$-yield are attributed to very small 
admixtures of $H$ contributions into clouds having
almost pure $P$-inventory composition. The rapid rise of 
$\log\epsilon({\rm Ba})$ at $-3\lesssim$~[Fe/H]~$<-2.5$
(between the two vertical dashed lines) is due to the onset of
$H$ events. The dot-dashed line shows the mean trend of data at
[Fe/H]~$>-2.5$ where both $H$ and $L$ events occur,
but cannot describe the data at lower [Fe/H]. Observed 
$\log\epsilon({\rm Ba})$ values at $-3\lesssim$~[Fe/H]~$<-2.5$
that correspond to $n_H\gtrsim 30$ $H$ events (e.g., the open circle
for CS 22892-052 and the linked crosses for CS 31082-001), as well as
those at [Fe/H]~$>-2.5$ that are far above the trend line,
may represent enrichments by surface contamination in binaries
instead of gross increases of Ba abundance in the ISM from which
the stars formed.\label{ba}}

\figcaption{The $P$-inventory and $H$ and $L$ yields 
(see Table \ref{phl}). Figure 2a shows the $P$-inventory (dot-dashed
curve), the $H$-yields (solid curve), and the standard solar 
$r$-component
(dotted curve) that is translated to coincide with the $H$-yields of
Ba to Au. Figure 2b shows the $L$-yields corresponding to the standard 
(short-dashed curve) and corrected (long-dashed curve) solar 
$r$-inventory and the $H$-yields (solid curve). The $L$-yields of Sr 
and Ba only exist for the corrected solar $r$-inventory of these two
elements. The $L$-contributions to elements above Ba are negligible.
\label{figphl}}

\figcaption{Data on [E/Eu]$_r\equiv\log({\rm E/Eu})-
\log({\rm E/Eu})_{\odot,r}$ (filled triangles: Gratton \& Sneden 1994;
open squares: Burris et al. 2000; asterisks: Johnson \& Bolte 2001)
shown as a function of [Fe/H] for Ba (a), La (b), Ce (c), Pr (d), 
Nd (e), Sm (f), Gd (g), and Dy (h).
Figure 3a shows a regular increase of Ba/Eu above the standard solar
$r$-process value of Arlandini et al. (1999) with increasing [Fe/H].
Figures 3b and 3h show that La/Eu and Dy/Eu are within 
$\approx 0.3$ dex of the standard solar $r$-process values (with
the exception of several data points). Figures 3c, 3d, 3f, and 3g 
show sparse data on Ce/Eu, Pr/Eu, Sm/Eu, and Gd/Eu with no 
indication of any wide spread from the standard solar $r$-process 
values. Figure 3e shows a wide scatter in Nd/Eu with no identifiable
trend. These data suggest that heavy $r$-elements from La and above
come only from $H$ events and that $s$-process contributions
are unimportant at [Fe/H]~$<-1$.\label{heavyeu}}

\figcaption{The difference between the abundances calculated
from the model and the data of Burris et al. (2000), 
$\Delta\log\epsilon({\rm E})\equiv
\log\epsilon_{\rm cal}({\rm E})-\log\epsilon_{\rm obs}({\rm E})$,
shown as a function of [Fe/H] for Sr (a), Y (b), Zr (c),
and Ba (d). Open squares are for the standard solar $r$-abundances 
(column 6 of Table \ref{phl}) and filled circles for the 
corrected values proposed here (column 9 of Table \ref{phl}). 
Filled squares indicate UMP stars with [Fe/H]~$\approx -3$.
We estimate the uncertainties in the calculated abundances to be
$\approx 0.2$~dex. These represent our generalized approximation 
of the uncertainties in the data for HD 115444 and CS 22892-052 
that are used to derive the model parameters. Burris et al. (2000) 
did not explicitly assign uncertainties to their data.
\label{figsyzb}}

\figcaption{The difference between the abundances calculated
from the model with the corrected solar $r$-inventory
and the data of Magain (1989; open diamonds) and Gratton \&
Sneden (1994; filled triangles), 
$\Delta\log\epsilon({\rm E})\equiv
\log\epsilon_{\rm cal}({\rm E})-\log\epsilon_{\rm obs}({\rm E})$,
shown as a function of [Fe/H] for Sr (a), Y (b), Zr (c),
and Ba (d). Note that by using the corrected solar $r$-inventory 
for these elements based on the data set of Burris et al. (2000),
the model describes the two different data sets shown in this
figure equally well. We estimate the uncertainties in the 
calculated abundances to be $\approx 0.2$~dex. These are 
comparable to the uncertainties in the data of Magain (1989)
and Gratton \& Sneden (1994). 
In Figure 5c, the data for the open diamonds 
at [Fe/H]~$=-2.75$ and $-2.43$ are taken from Zhao \& Magain 
(1991).\label{test}}

\clearpage
\begin{table}
\begin{center}
\caption{Chronology of evolution\label{epoch}}
\begin{tabular}{cl}\tableline\tableline
Epoch&\multicolumn{1}{c}{Description}\\ \tableline 
(1)&Big Bang\\
(2)& Formation and explosion of very massive ($\gtrsim
100\,M_\odot$) stars\\ 
&in regions of baryonic matter condensation
with production of ``metals'' \\ 
&(C, N, O, Mg, Al, Si, ..., Fe, ..., Sr, Y, Zr)\\ 
(3)&Disruption of condensed regions by explosions
of very massive stars,\\ 
&return of material with ``metals''
to the IGM, and reionization of the IGM\\ 
(4)&Cooling and
condensation of some of the baryonic matter in the IGM\\
(5)&Formation of normal stellar populations with masses of $\sim
(1$--$60)\,M_\odot$\\ 
&when a metallicity of $-4<{\rm
[Fe/H]}<-2.7$ was reached\\ 
(6)&Occurrence of the first SNII $H$
events [with a frequency of $\sim (10^7\  {\rm yr})^{-1}$\\ 
&for a standard reference mass of hydrogen]\\ 
(7)&Coeval occurrence of
SNII $H$ and $L$ events [with an $L$-frequency of\\ 
&$\sim (10^8\
{\rm yr})^{-1}$ for a standard reference mass of hydrogen]\\
(8)&Later occurrence of SNIa with major Fe
addition and evolution of low-mass AGB\\ 
&stars resulting in substantial $s$-process contributions 
to the ISM at [Fe/H]~$\gtrsim -1$\\ \tableline
\end{tabular}
\end{center}
\end{table}

\clearpage
\begin{deluxetable}{crrrrrrrr}
\tabletypesize{\footnotesize}
\tablecaption{Data on three UMP stars and calculated
numbers of $H$ and $L$ events\label{syzph}}
\tablewidth{0pt}
\tablehead{
\colhead{Star}&\colhead{[Fe/H]}&
\colhead{$n_L$}&
\colhead{$\log\epsilon({\rm Eu})$}&
\colhead{$n_H$}&
\colhead{$\log\epsilon({\rm Sr})$}&\colhead{$\log\epsilon({\rm Y})$}&
\colhead{$\log\epsilon({\rm Zr})$}&
\colhead{Ref.}
}
\startdata
HD 122563&$-2.74$&0&$-2.59$&1&0.33&$-0.75$&0.04&[1]\\
HD 115444&$-2.99$&0&$-1.63$&7&0.23&$-0.82$&$-0.02$&[1]\\
CS 22892-052&$-3.1$&0&$-0.93$&36&0.50&$-0.39$&0.26&[2]\\
\enddata
\tablerefs{[1] Westin et al. 2000;
[2] Sneden et al. 2000.}
\end{deluxetable}

\clearpage
\begin{deluxetable}{ccrrrrrrrr}
\tabletypesize{\footnotesize}
\tablecaption{Parameters for the three-component model\label{phl}}
\tablewidth{0pt}
\tablehead{\colhead{Z\tablenotemark{a}}&
\colhead{E}&\colhead{$\log\epsilon_P({\rm E})$}&
\colhead{$\log\epsilon_H({\rm E})$}&
\colhead{$\log\epsilon_L({\rm E})$}&
\colhead{$\log\epsilon_{\odot,r}({\rm E})$}&
\colhead{$\beta_{\odot,s}({\rm E})$}&
\colhead{$\log\epsilon_L^{\rm corr}({\rm E})$}&
\colhead{$\log\epsilon_{\odot,r}^{\rm corr}({\rm E})$}&
\colhead{$\beta_{\odot,s}^{\rm corr}({\rm E})$}
}
\startdata
26&Fe\tablenotemark{b}&
4.51&$-\infty$&5.03&\nodata&\nodata&\nodata&\nodata&\nodata\\
38&Sr&0.13&$-1.30$&$-\infty$\tablenotemark{c}&
1.64&0.95\tablenotemark{d}&0.35&2.44&0.69\\
39&Y&$-1.05$&$-2.05$&$-1.36$&1.12&0.92&$-0.42$&1.67&0.72\\
40&Zr&$-0.13$&$-1.53$&$-0.40$&1.85&0.83&0.26&2.32&0.49\\
41&Nb&$-\infty$&$-2.56$&$-2.06$&0.56&0.85&\nodata&\nodata&\nodata\\
44&Ru&$-\infty$&$-1.56$&$-0.92$&1.60&0.32&\nodata&\nodata&\nodata\\
45&Rh&$-\infty$&$-2.36$&$-1.20$&1.03&0.14&\nodata&\nodata&\nodata\\
46&Pd&$-\infty$&$-1.82$&$-0.95$&1.42&0.46&\nodata&\nodata&\nodata\\
47&Ag&$-\infty$&$-2.36$&$-1.02$&1.14&0.20&\nodata&\nodata&\nodata\\
48&Cd&$-\infty$&$-1.91$&$-0.85$&1.42&0.52&\nodata&\nodata&\nodata\\
56&Ba&$-\infty$&$-1.52$\tablenotemark{e}&
$-\infty$&1.48&0.81&$-0.47$&1.79&0.62\\
57&La&$-\infty$&$-2.22$&$-\infty$&0.78&0.62&\nodata&\nodata&\nodata\\
58&Ce&$-\infty$&$-2.02$&$-\infty$&0.98&0.77&\nodata&\nodata&\nodata\\
59&Pr&$-\infty$&$-2.51$&$-\infty$&0.49&0.49&\nodata&\nodata&\nodata\\
60&Nd&$-\infty$&$-1.90$&$-\infty$&1.10&0.56&\nodata&\nodata&\nodata\\
62&Sm&$-\infty$&$-2.20$&$-\infty$&0.80&0.29&\nodata&\nodata&\nodata\\
63&Eu&$-\infty$&$-2.48$&$-\infty$&0.52&0.06&\nodata&\nodata&\nodata\\
64&Gd&$-\infty$&$-2.00$&$-\infty$&1.00&0.15&\nodata&\nodata&\nodata\\
65&Tb&$-\infty$&$-2.70$&$-\infty$&0.30&0.07&\nodata&\nodata&\nodata\\
66&Dy&$-\infty$&$-1.92$&$-\infty$&1.08&0.15&\nodata&\nodata&\nodata\\
67&Ho&$-\infty$&$-2.53$&$-\infty$&0.47&0.08&\nodata&\nodata&\nodata\\
68&Er&$-\infty$&$-2.13$&$-\infty$&0.87&0.17&\nodata&\nodata&\nodata\\
69&Tm&$-\infty$&$-2.93$&$-\infty$&0.07&0.13&\nodata&\nodata&\nodata\\
70&Yb&$-\infty$&$-2.22$&$-\infty$&0.78&0.33&\nodata&\nodata&\nodata\\
71&Lu&$-\infty$&$-2.98$&$-\infty$&0.02&0.20&\nodata&\nodata&\nodata\\
72&Hf&$-\infty$&$-2.61$&$-\infty$&0.39&0.56&\nodata&\nodata&\nodata\\
73&Ta&$-\infty$&$-3.36$&$-\infty$&$-0.36$&0.41&\nodata&\nodata&\nodata\\
74&W&$-\infty$&$-2.68$&$-\infty$&0.32&0.56&\nodata&\nodata&\nodata\\
75&Re&$-\infty$&$-2.75$&$-\infty$&0.25&0.09&\nodata&\nodata&\nodata\\
76&Os&$-\infty$&$-1.66$&$-\infty$&1.34&0.09&\nodata&\nodata&\nodata\\
77&Ir&$-\infty$&$-1.63$&$-\infty$&1.37&0.01&\nodata&\nodata&\nodata\\
78&Pt&$-\infty$&$-1.34$&$-\infty$&1.66&0.05&\nodata&\nodata&\nodata\\
79&Au&$-\infty$&$-2.20$&$-\infty$&0.80&0.06&\nodata&\nodata&\nodata\\
90&$^{232}$Th\tablenotemark{f}&$-\infty$&$-2.72$&$-\infty$&0.18&0&
\nodata&\nodata&\nodata\\
92&$^{238}$U\tablenotemark{f}&$-\infty$&$-2.89$&$-\infty$&$-0.19$&0&
\nodata&\nodata&\nodata\\
\enddata
\tablenotetext{a}{Atomic number.}
\tablenotetext{b}{Use $\log\epsilon_\odot({\rm Fe})=7.51$ to obtain
[Fe/H]~$\equiv\log\epsilon({\rm Fe})-\log\epsilon_\odot({\rm Fe})$.
The $P$-inventory and $H$ and $L$ yields of Fe are inferred from
the data on UMP stars and the part of the solar Fe inventory
contributed by SNII.}
\tablenotetext{c}{The standard solar $r$-inventory of Sr 
is saturated by the contributions 
from $n_H^\odot=10^3$ $H$ events and thus does not allow any $L$ 
contributions to Sr.}
\tablenotetext{d}{This $\beta_{\odot,s}({\rm Sr})$ value is based on
the solar $r$-process fraction of
$[1-\beta_{\odot,s}({\rm Sr})]=0.05$ given by Arlandini et al. 1999.}
\tablenotetext{e}{This $H$-yield of Ba is calculated by attributing
the standard solar $r$-inventory of Ba to $n_H^\odot=10^3$ $H$ events
and is essentially identical to the value of 
$\log\epsilon_H({\rm Ba})\approx -1.57$ obtained from the data on
CS 22892-052. The latter value is used in the calculations.}
\tablenotetext{f}{The $H$-yields of $^{232}$Th and $^{238}$U
are taken from the calculations of QW based on the inventory
of these two nuclei in the early solar system (given here as
the $\log\epsilon_{\odot,r}$ values).}
\tablecomments{The $P$-inventory in column 3 and $H$-yields in
column 4 are calculated from the data on HD 115444 and CS 22892-052
in Table 1 for Sr, Y, and Zr.
The $P$-inventory for elements above Zr is negligible and set
to $-\infty$. The other $H$-yields are calculated from
the data on CS 22892-052 for Nb to Cd and from the standard
solar $r$-abundances (Arlandini et al. 1999)
in column 6 (corresponding to the solar
$s$-process fractions in column 7) for Ba to Au. The $L$-yields 
in column 5 are calculated from columns 3, 4, and 6 for Sr to Cd
and set to $-\infty$ for Ba and above. The last three columns
give the corrected values proposed here for Sr, Y, Zr, and Ba.}
\end{deluxetable}


\clearpage
\begin{references}
\reference{}
Anders, E., \& Grevesse, N. 1989, \gca, 53, 197

\reference{} 
Aoki, W., Norris, J. E., Ryan, S. G., Beers, T. C.,
\& Ando, H. 2000, \apjl, 536, L97

\reference{}
Arlandini, C., K\"appler, F., Wisshak, K., Gallino, R., Lugaro, M.,
Busso, M., \& Straniero, O. 1999, \apj, 525, 886 

\reference{}
Baraffe, I., Heger, A., \& Woosley, S. E. 2001, \apj, 550, 890

\reference{}
Brazzle, R. H., Pravdivtseva, O. V., Meshik, A. P., \& 
Hohenberg, C. M. 1999, \gca, 63, 739

\reference{}
Bromm, V., Coppi, P. S., \& Larson, R. B. 2000, \apjl, 527, L5

\reference{}
Bromm, V., Ferrara, A., Coppi, P. S., \& Larson, R. B. 2001,
\mnras, submitted

\reference{}
Burbidge, E. M., Burbidge, G. R., Fowler, W. A., \& Hoyle, F. 1957,
Rev. Mod. Phys., 29, 547

\reference{} 
Burris, D. L., Pilachowski, C. A., Armandroff, T. E.,
Sneden, C., Cowan, J. J., \& Roe, H. 2000, \apj, 544, 302 

\reference{}
Busso, M., Gallino, R., \& Wasserburg, G. J. 1999, \araa, 37, 239

\reference{}
Cameron, A. G. W. 1957, \pasp, 69, 201

\reference{}
Cayrel, R., et al. 2001, \nat, 409, 691

\reference{}
Ezer, D., \& Cameron, A. G. W. 1971, \apss, 14, 399 

\reference{}
Freiburghaus, C., Rembges, J.-F., Rauscher, T., Kolbe, E., 
Thielemann, F.-K.,
Kratz, K.-L., Pfeiffer, B., \& Cowan, J. J. 1999a, \apj, 516, 381

\reference{}
Fryer, C. L., Woosley, S. E., \& Heger, A. 2001, \apj, 550, 372

\reference{}
Gallino, R., Arlandini, C., Busso, M., Lugaro, M., Travaglio, C.,
Straniero, O., Chieffi, A., \& Limongi, M. 1998, \apj, 497, 388

\reference{}
Gratton, R. G., \& Sneden, C. 1994, \aap, 287, 927

\reference{}
Heger, A., Baraffe, I., Fryer, C. L., \& Woosley, S. E. 2000,
astro-ph/0010206

\reference{}
Hill, V., Plez, B., Cayrel, R., \& Beers, T. C. 2001,
astro-ph/0104172

\reference{}
Hoffman, R. D., Woosley, S. E., \& Qian, Y.-Z. 1997, \apj, 482, 951

\reference{}
Johnson, J. A., \& Bolte, M. 2001, \apj, in press

\reference{}
K\"appler, F., Beer, H. \& Wisshak, K. 1989, 
Rep. Prog. Phys., 52, 945

\reference{}
Kratz, K.-L., Bitouzet, J.-P., Thielemann, F.-K., M\"oller, P., \& 
Pfeiffer, B. 1993, \apj, 403, 216

\reference{}
Magain, P. 1989, \aap, 209, 211

\reference{}
McWilliam, A., Preston, G. W., Sneden, C., \& Searle, L. 1995,
\aj, 109, 2757

\reference{} 
McWilliam, A. 1998, \aj, 115, 1640

\reference{}
Meyer, B. S., \& Brown, J. S. 1997, \apjs, 112, 199

\reference{}
Preston, G. W., \& Sneden, C. 2001, \apj, submitted

\reference{}
Qian, Y.-Z. 2000, \apjl, 534, L67

\reference{}
Qian, Y.-Z. 2001, \apjl, 552, L117

\reference{}
Qian, Y.-Z., Vogel, P., \& Wasserburg, G. J., 1998,
\apj, 494, 285

\reference{}
Qian, Y.-Z., \& Wasserburg, G. J. 2000, Phys. Rep., 333--334, 77 

\reference{} 
Qian, Y.-Z., \& Wasserburg, G. J. 2001a, \apj, 549, 337

\reference{} 
Qian, Y.-Z., \& Wasserburg, G. J. 2001b, \apjl, 552, L55 (QW)

\reference{} 
Raiteri, C. M., Villata, M., Gallino, R., Busso, M., 
\& Cravanzola, A. 1999, \apj, 518, L91
 
\reference{} 
Ryan, S. G., Norris, J. E., \& Beers, T. C. 1996, \apj, 471, 254 

\reference{}
Sneden,C.,McWilliam, A., Preston, G. W., Cowan, J. J., Burris, D.
L., \& Armosky, B. J. 1996, \apj, 467, 819 

\reference{} 
Sneden, C., Cowan, J. J., Ivans, I. I., Fuller, G. M., Burles, S., 
Beers, T. C., \& Lawler, J. E. 2000, \apj, 533, L139 

\reference{}
Thornton, K., Gaudlitz, M., Janka, H.-Th., \& Steinmetz, M. 1998,
\apj, 500, 95

\reference{} 
Timmes, F. X., Woosley, S. E., \& Weaver, T. A. 1995, 
\apjs, 98, 617

\reference{} 
Travaglio, C., Galli, D., Gallino, R., Busso, M.,
Ferrini, F., \& Straniero, O. 1999, \apj, 521, 691 

\reference{}
Truran, J. W. 1981, \aap, 97, 391

\reference{}
Wasserburg, G. J., Busso, M., \& Gallino, R. 1996, \apj, 466, L109
(WBG) 

\reference{} 
Wasserburg, G. J., \& Qian, Y.-Z. 2000, \apjl, 529, L21 (WQ) 

\reference{} 
Westin, J., Sneden, C., Gustafsson, B.,
\& Cowan, J. J. 2000, \apj, 530, 783 

\reference{} 
Woosley, S. E., \& Hoffman, R. D. 1992, \apj, 395, 202

\reference{}
Zhao, G., \& Magain, P. 1991, \aap, 244, 425
\end{references}
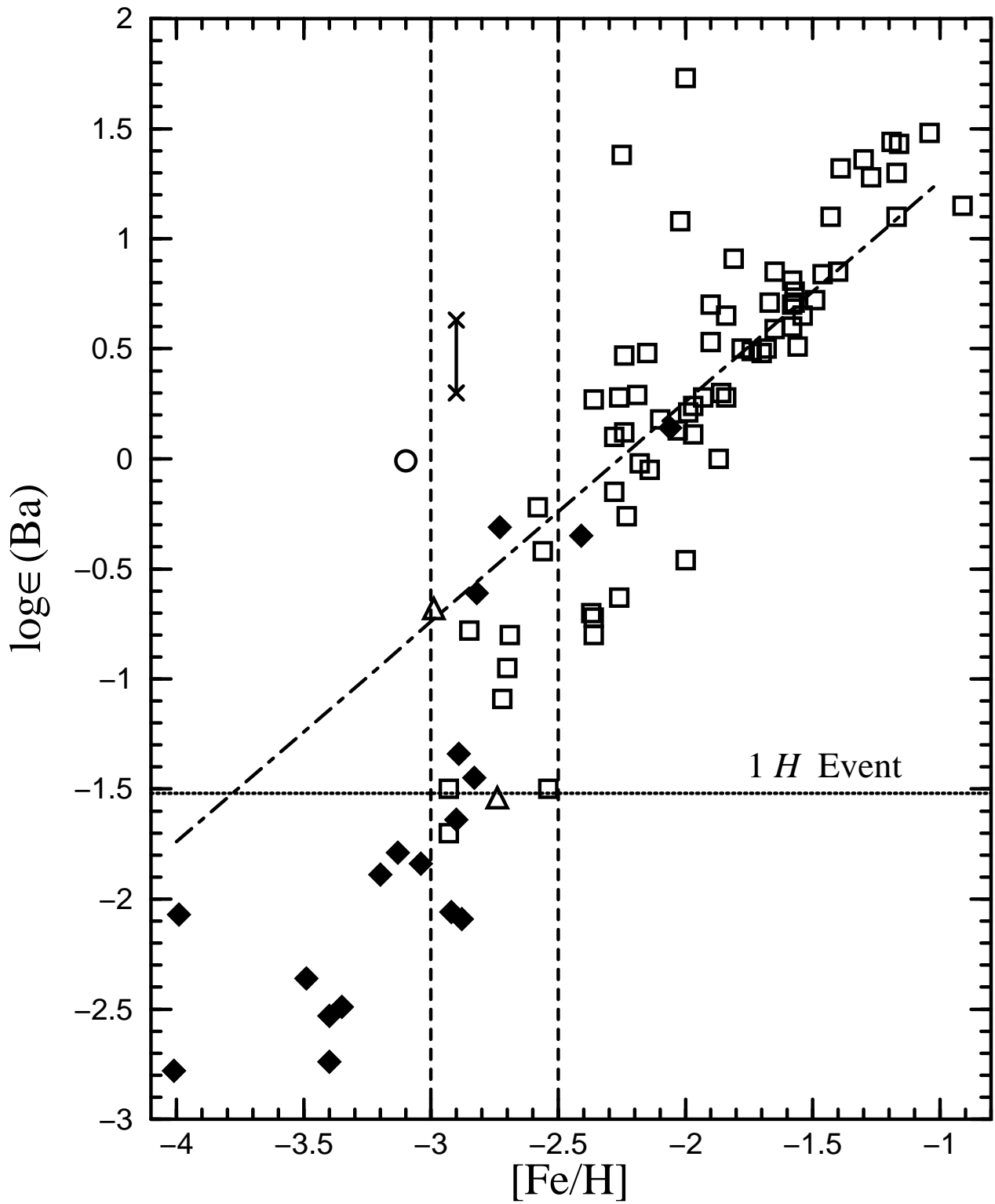
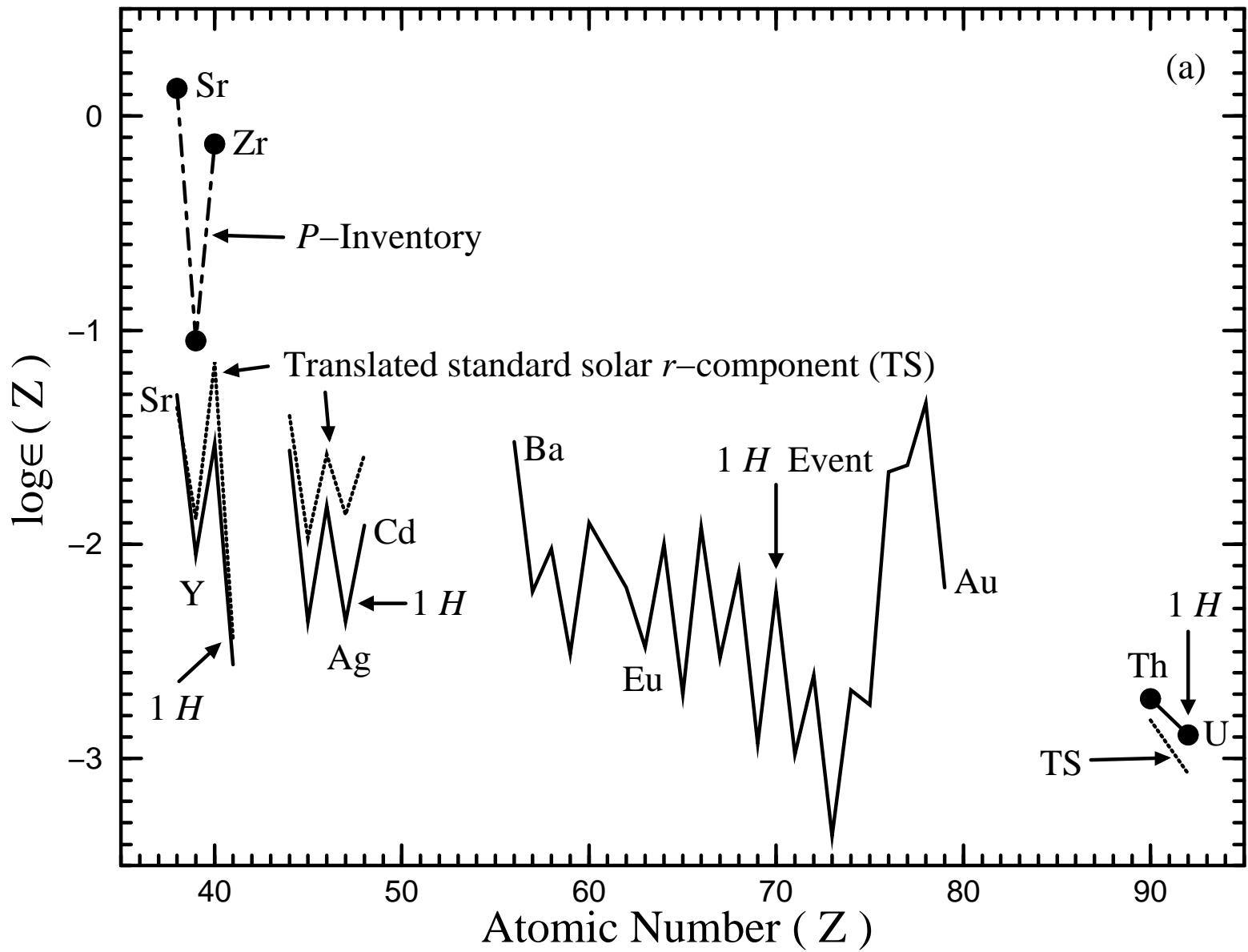
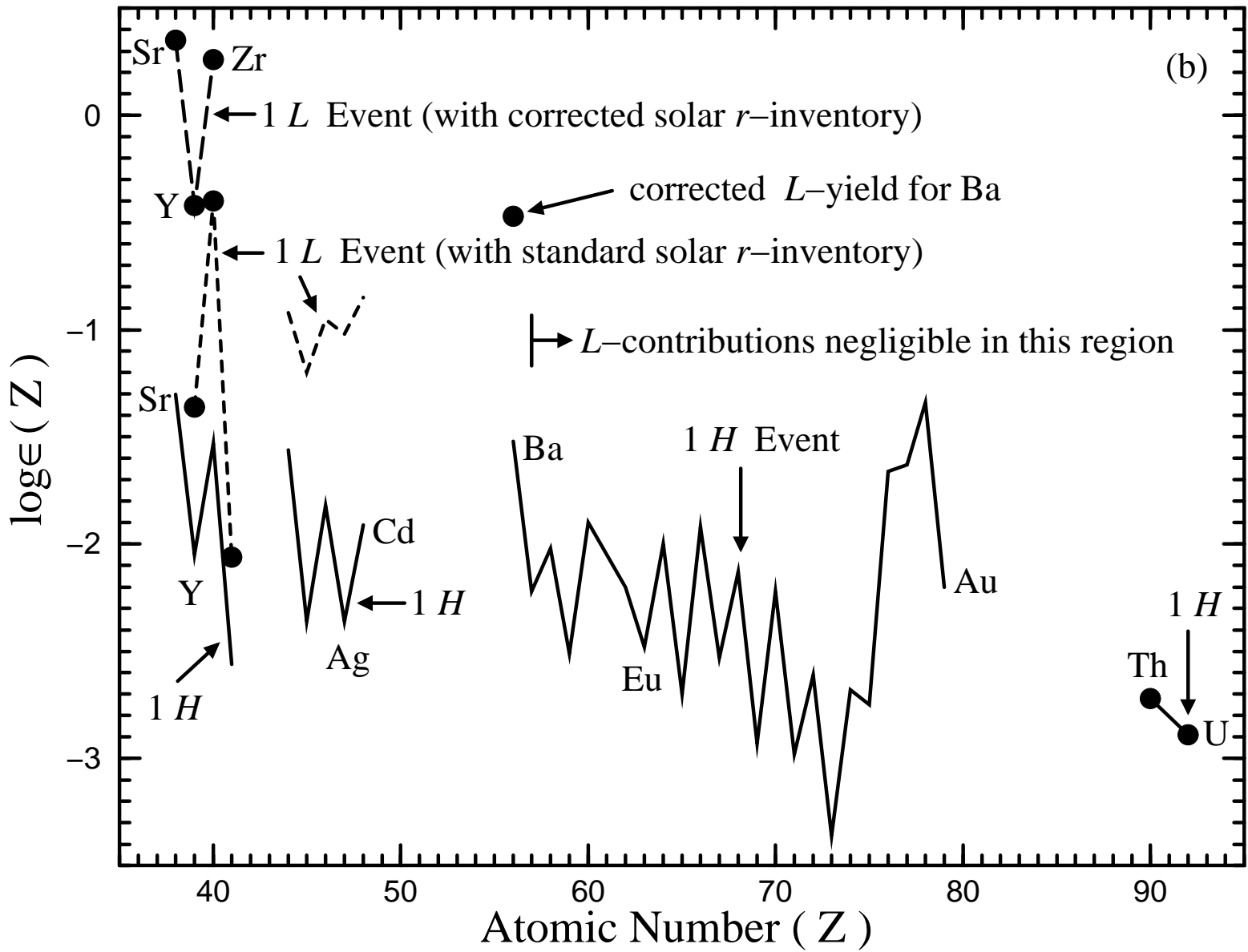
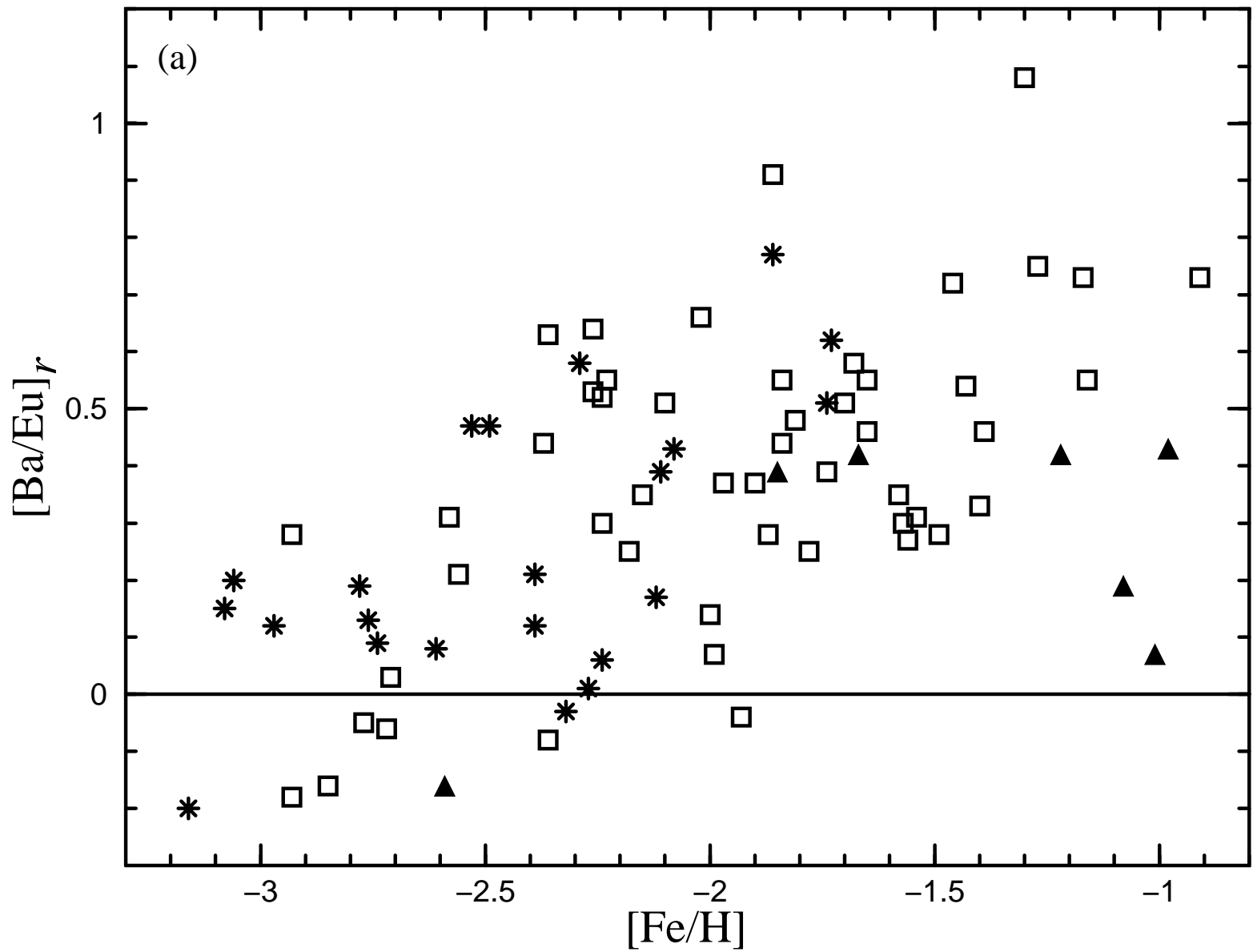
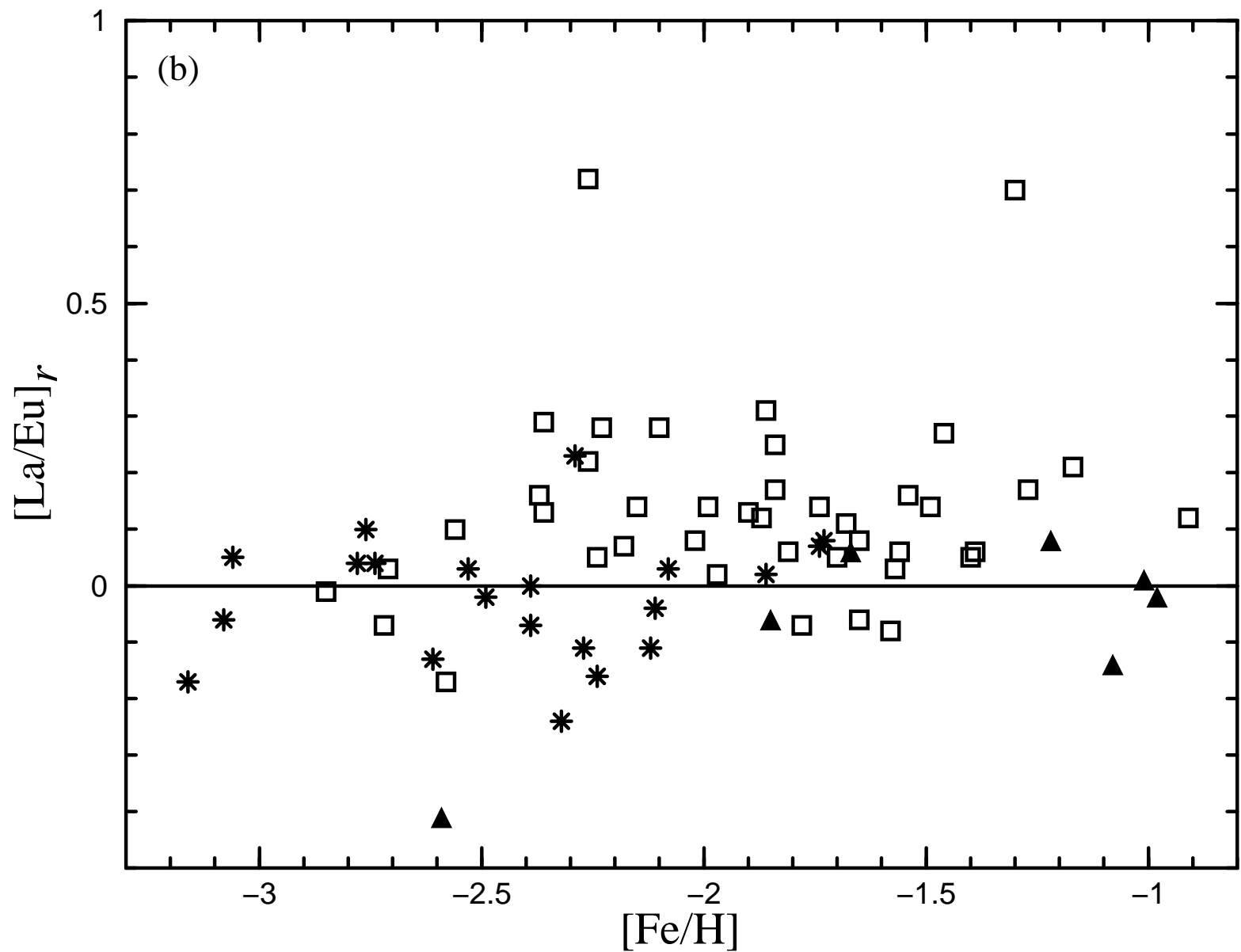
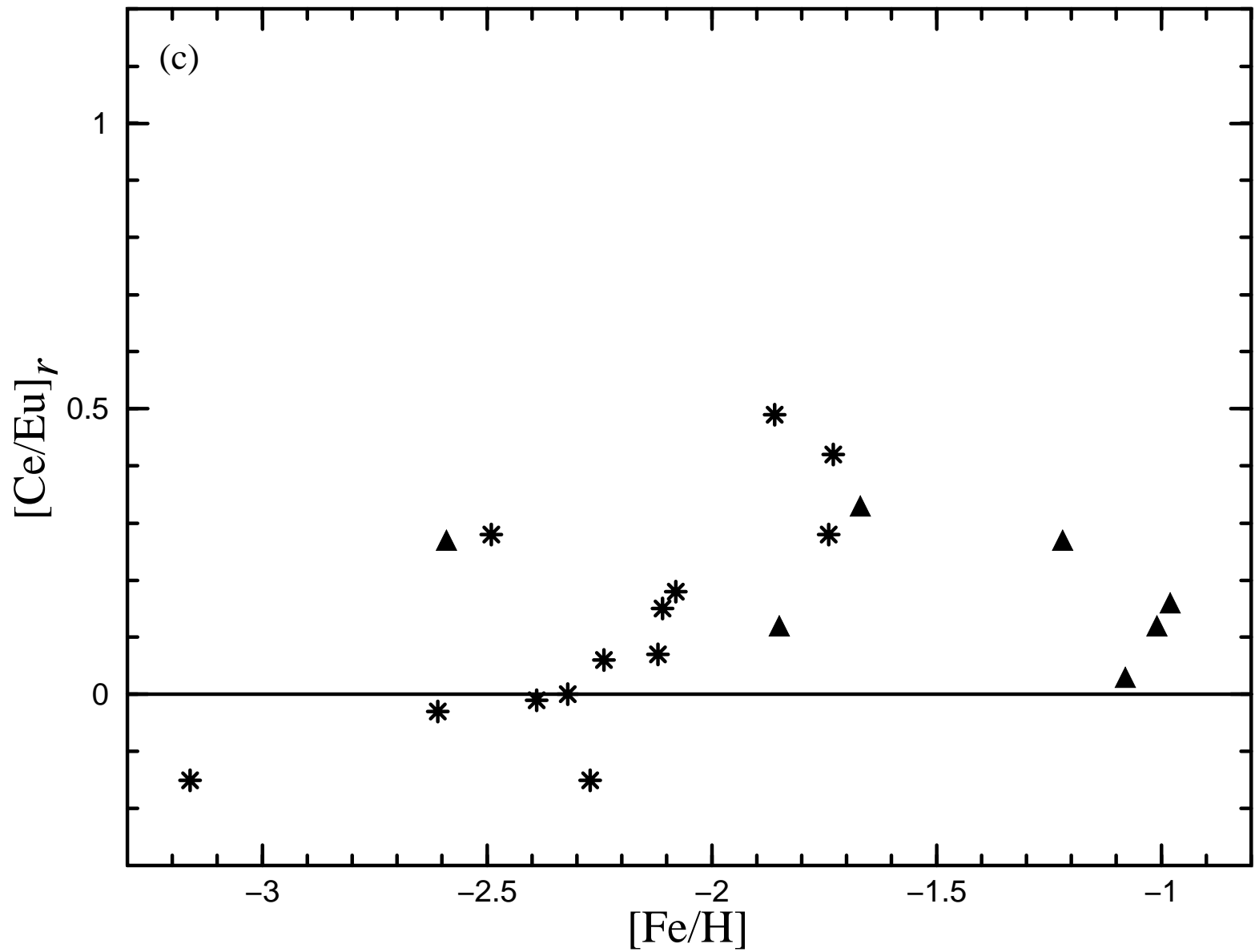
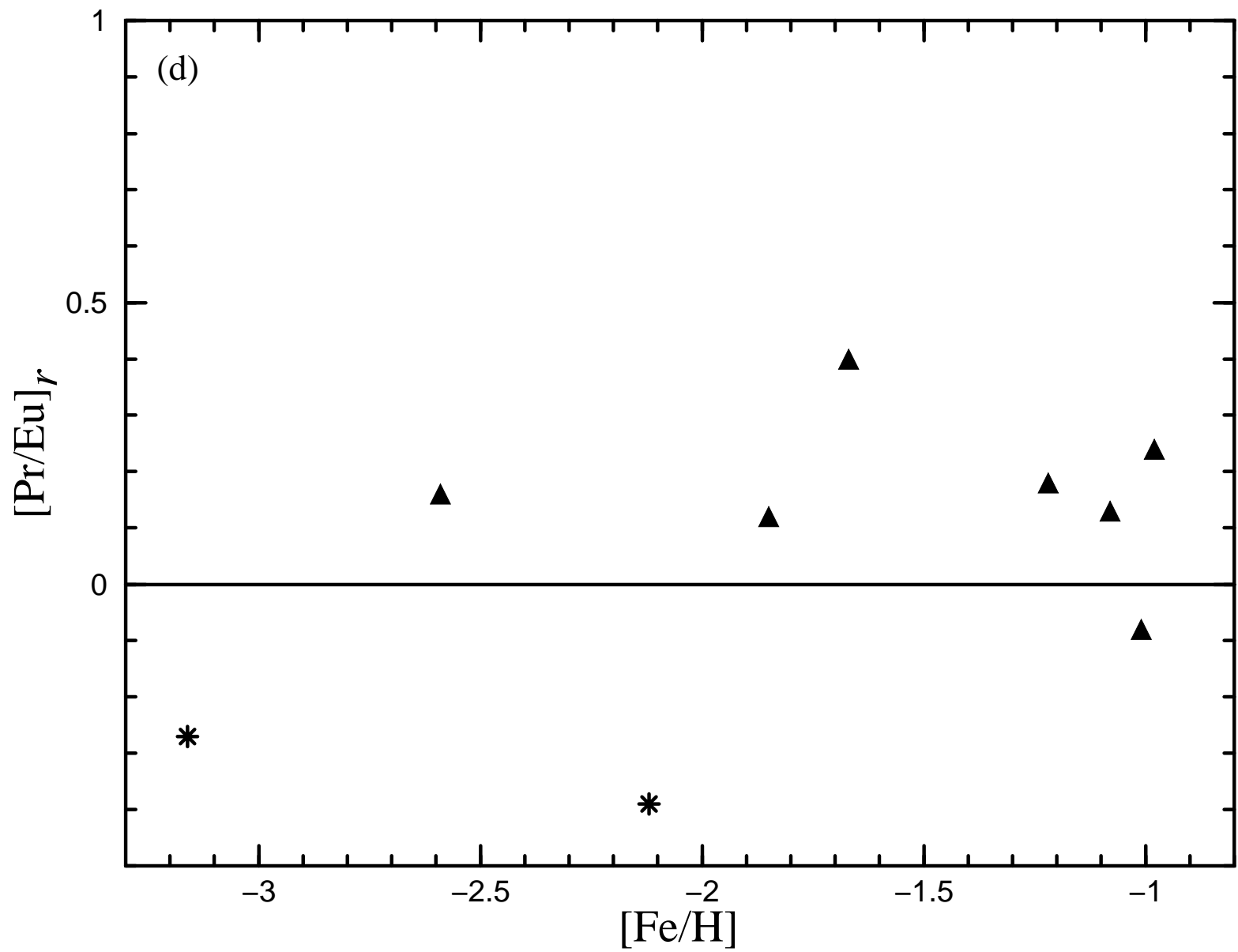
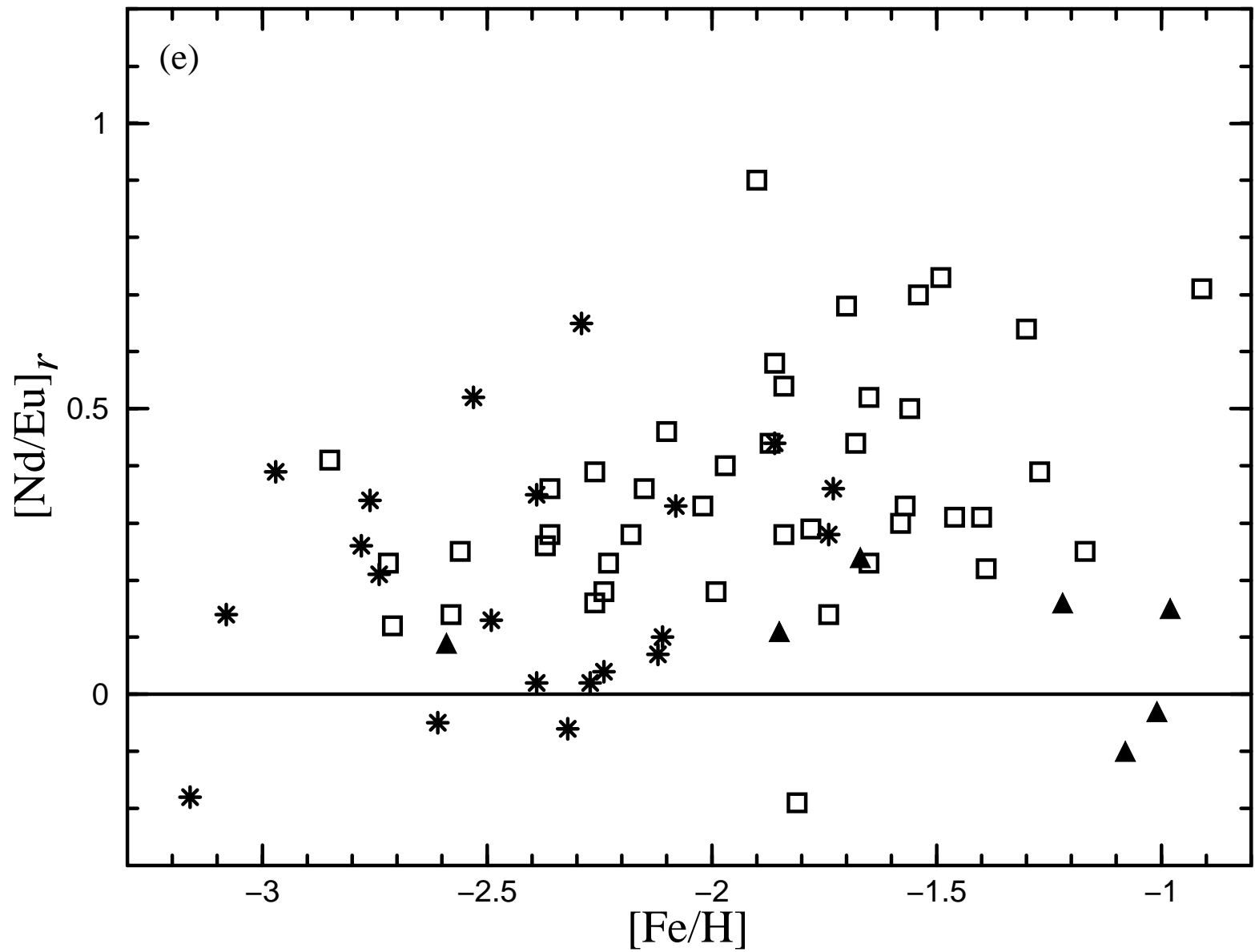
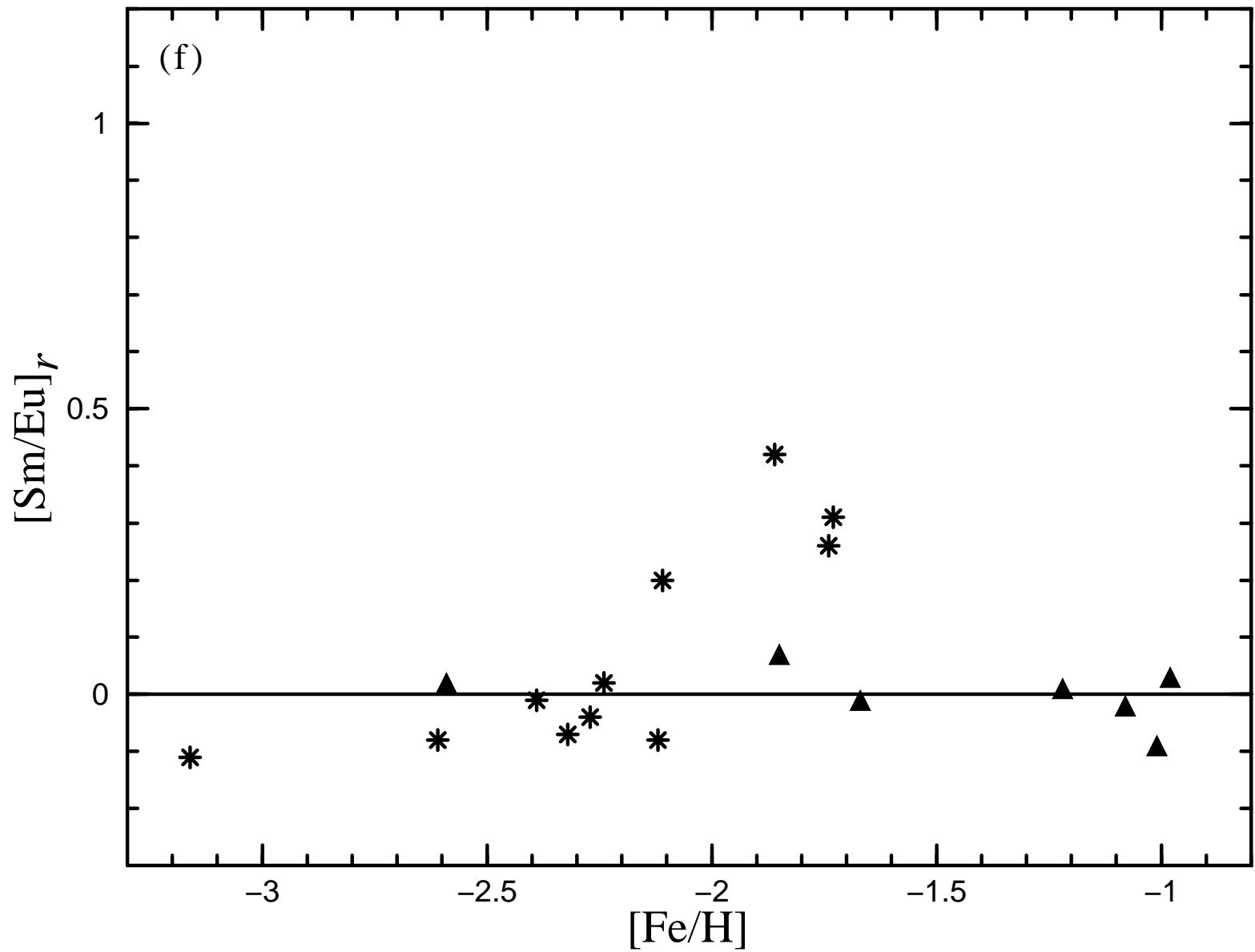
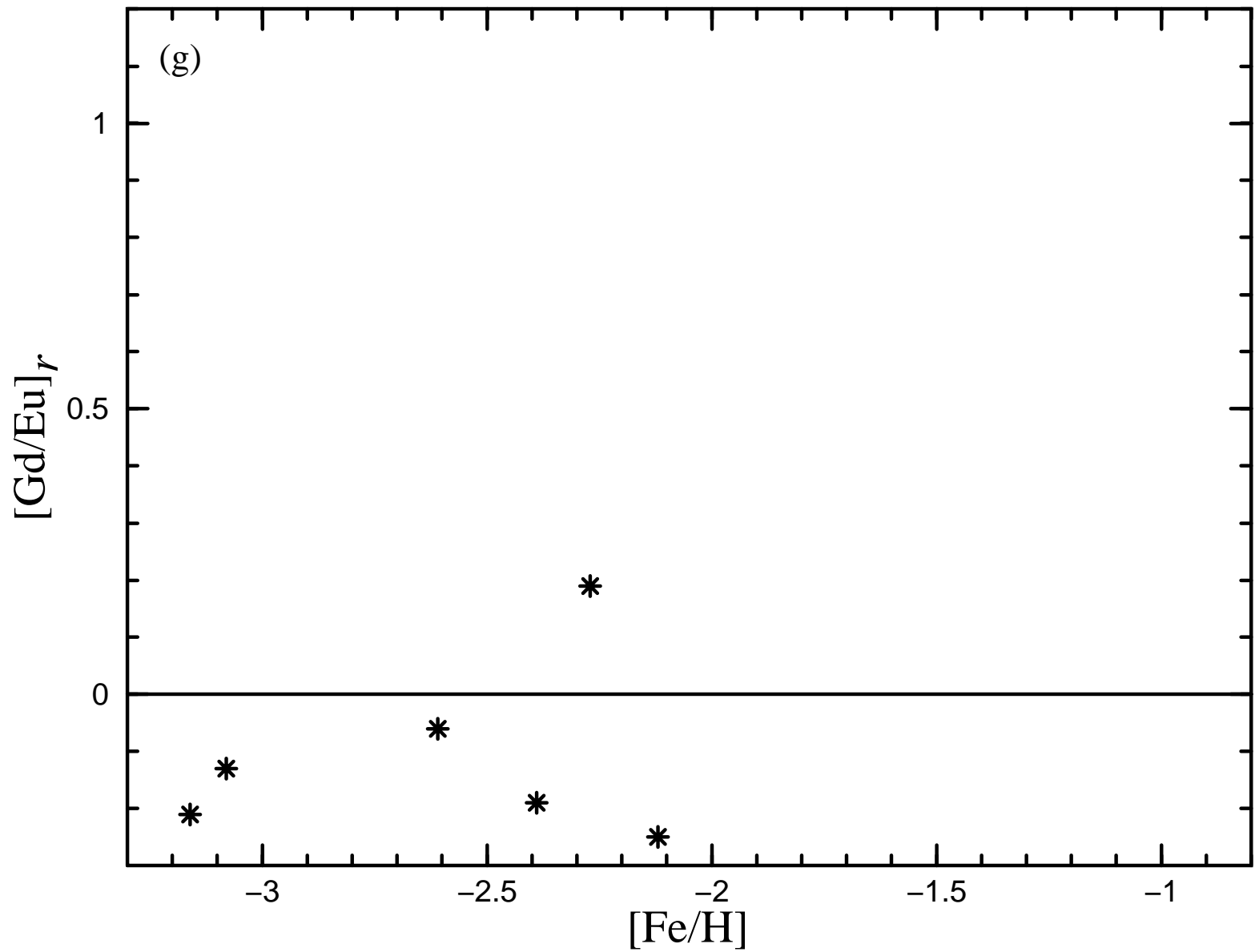
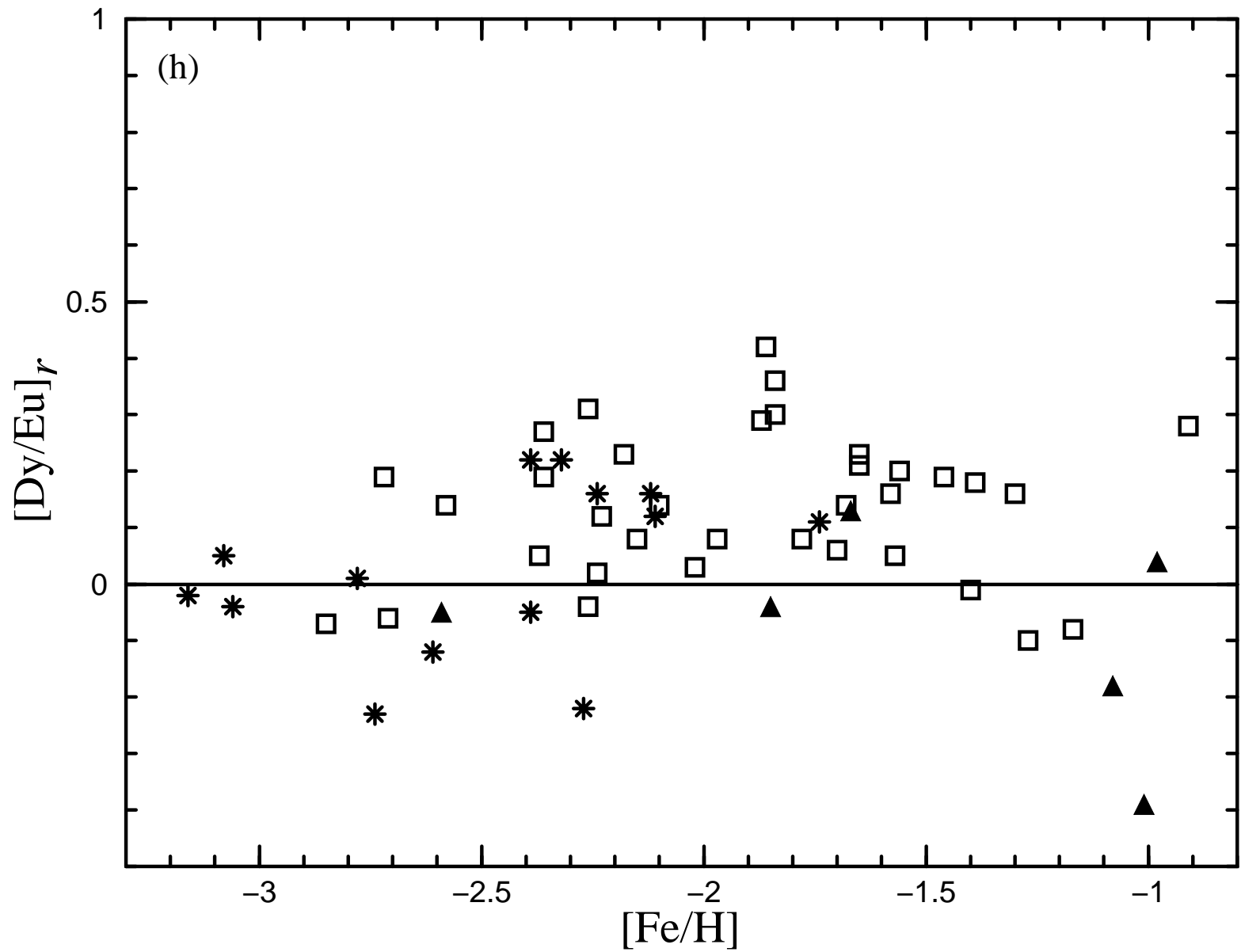
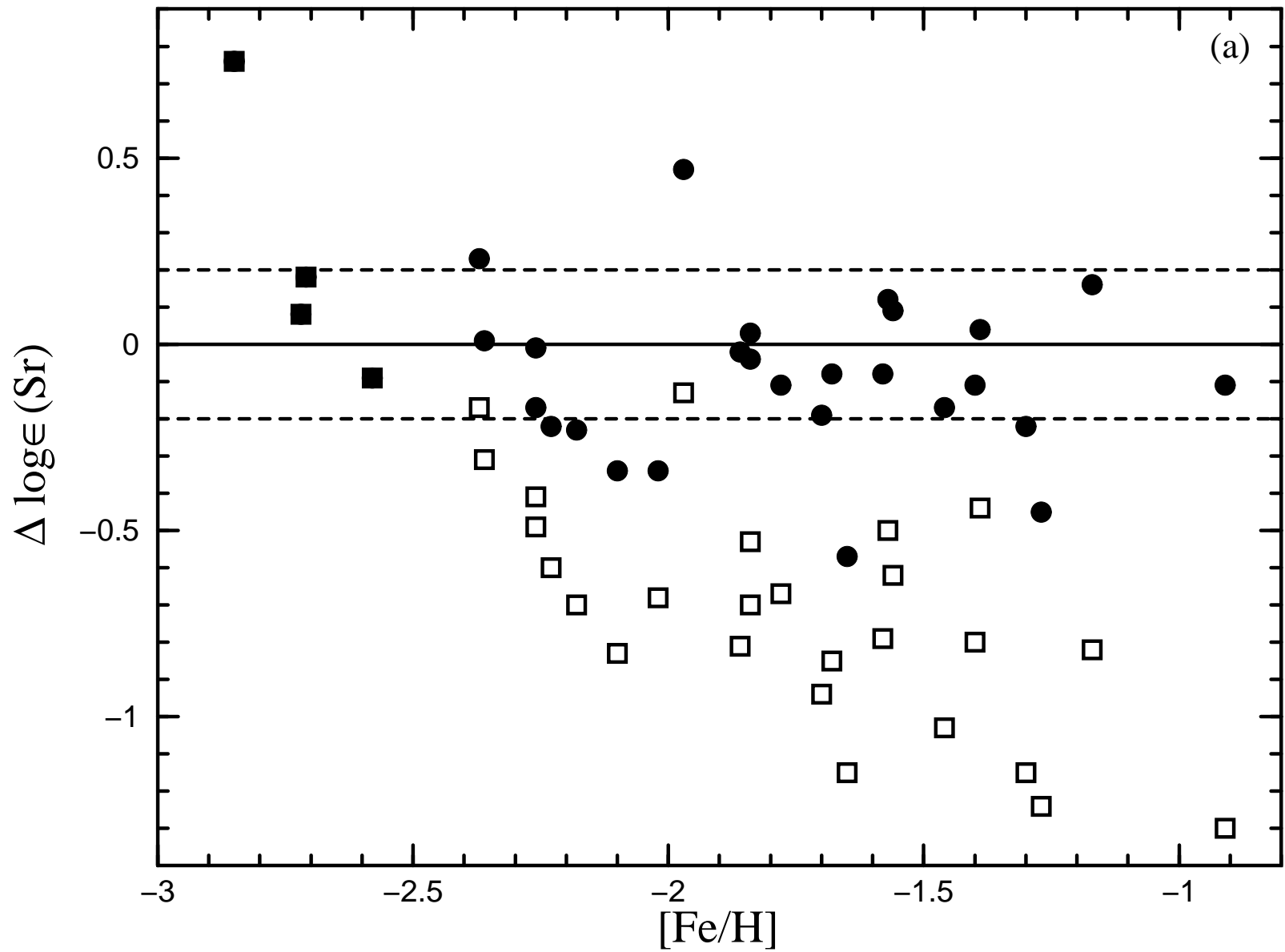
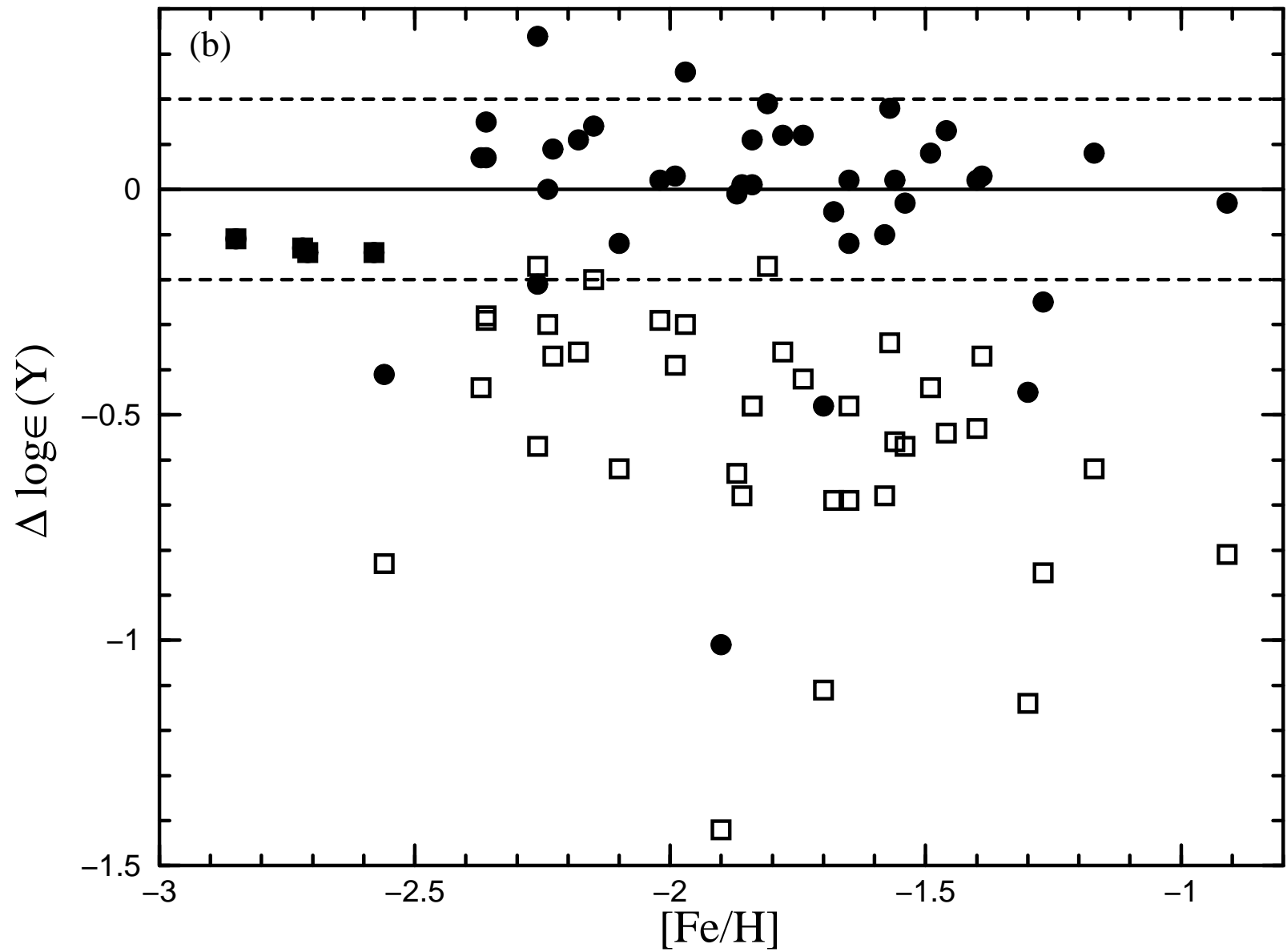
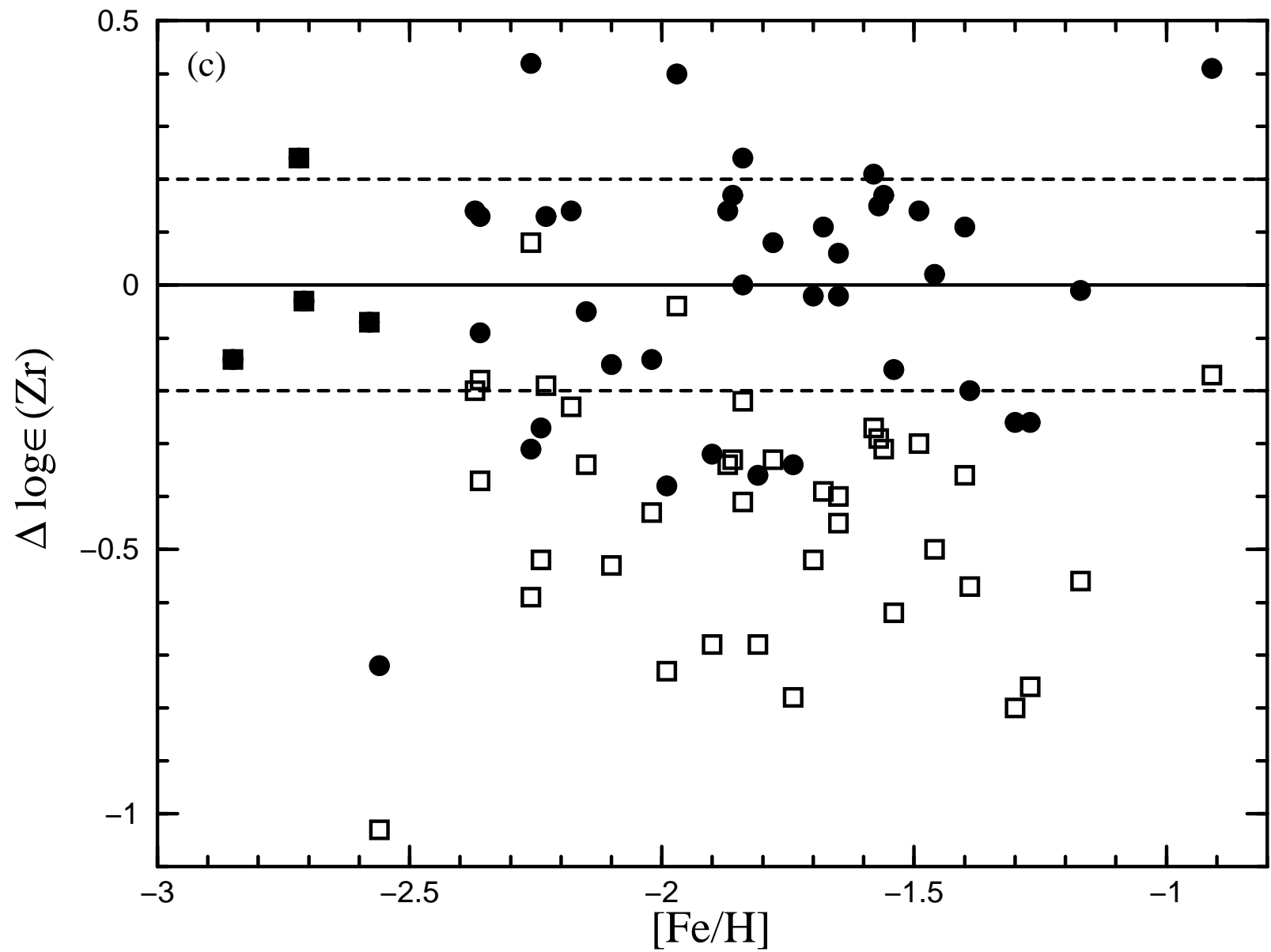
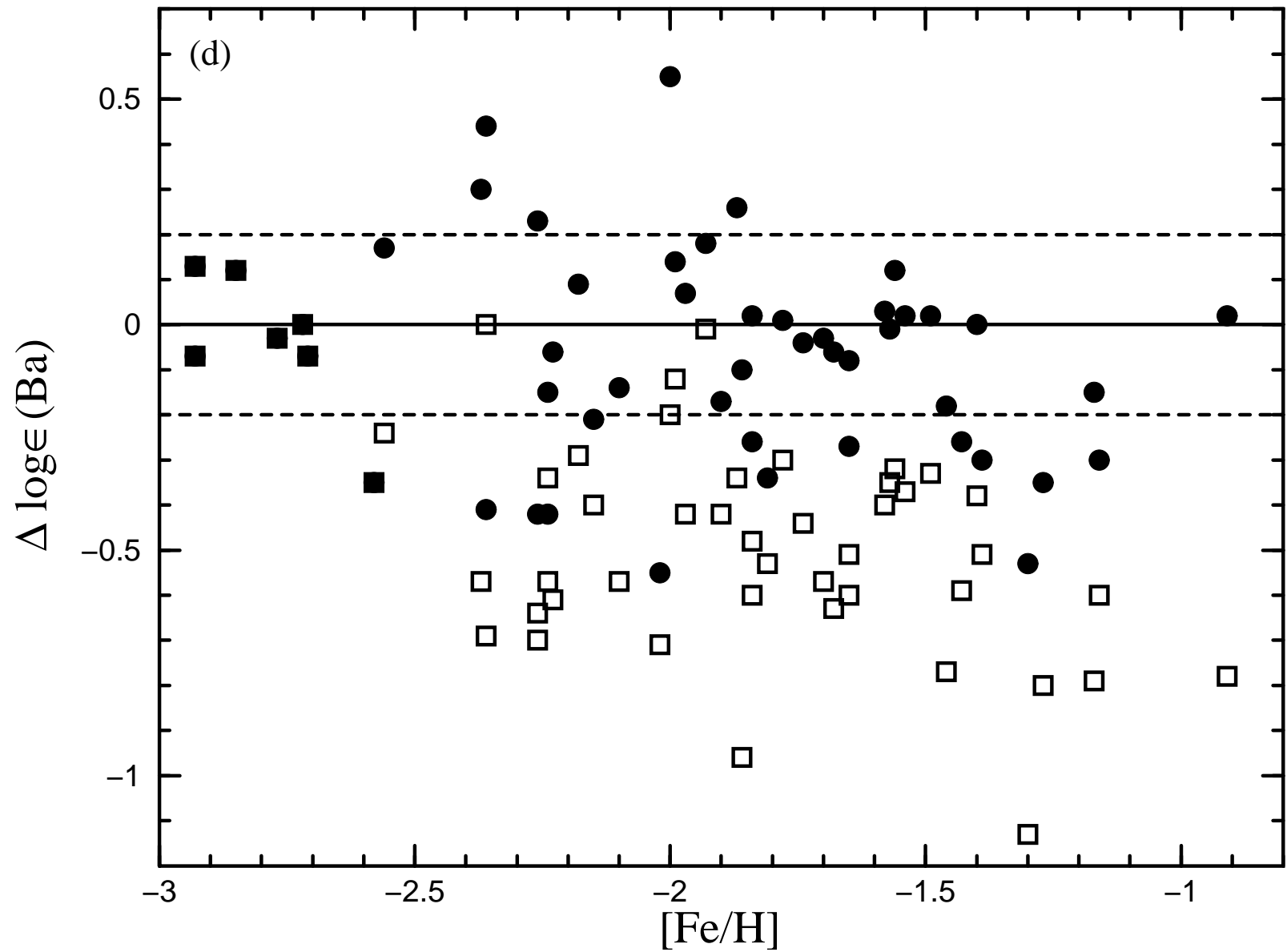
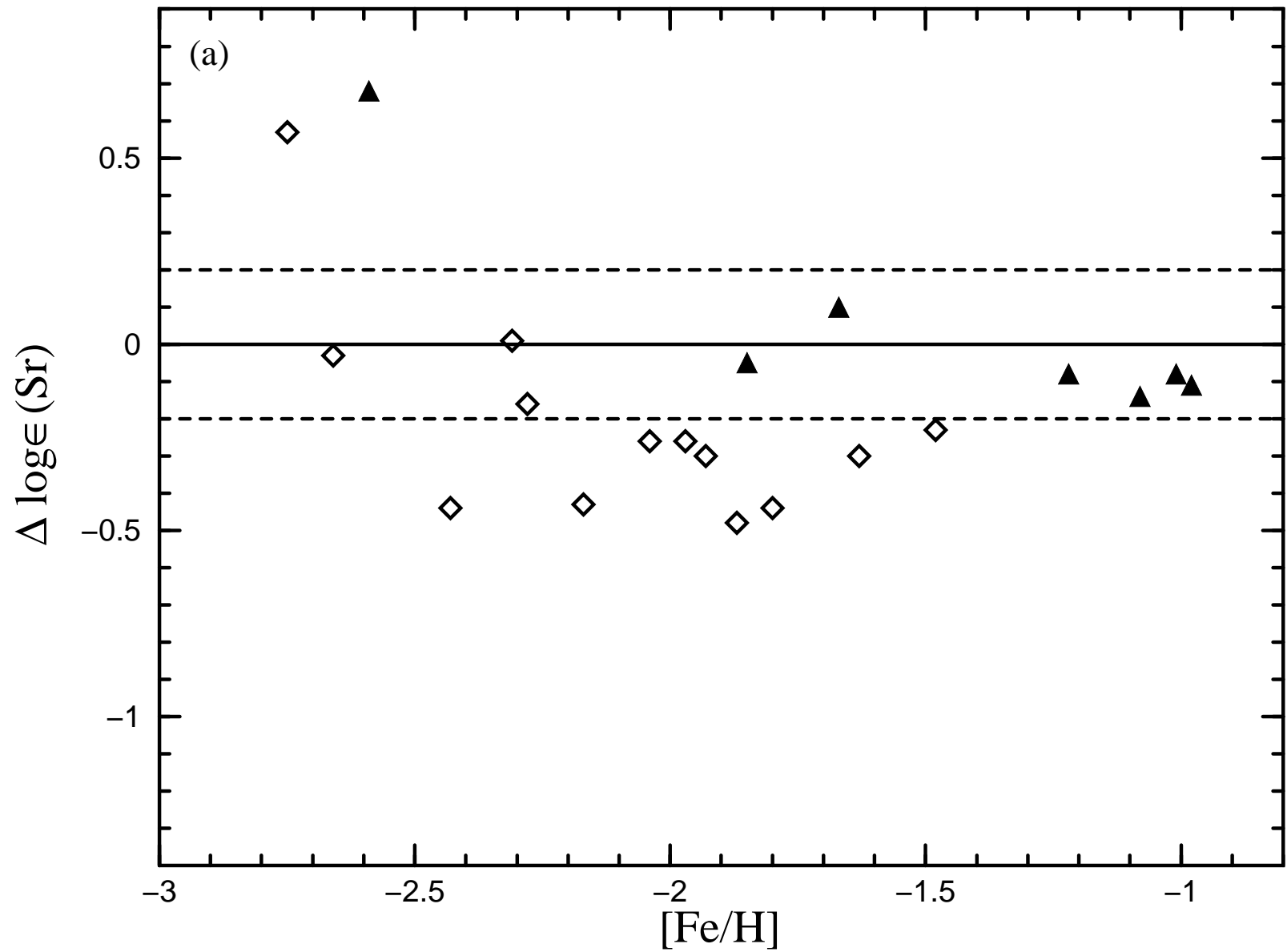
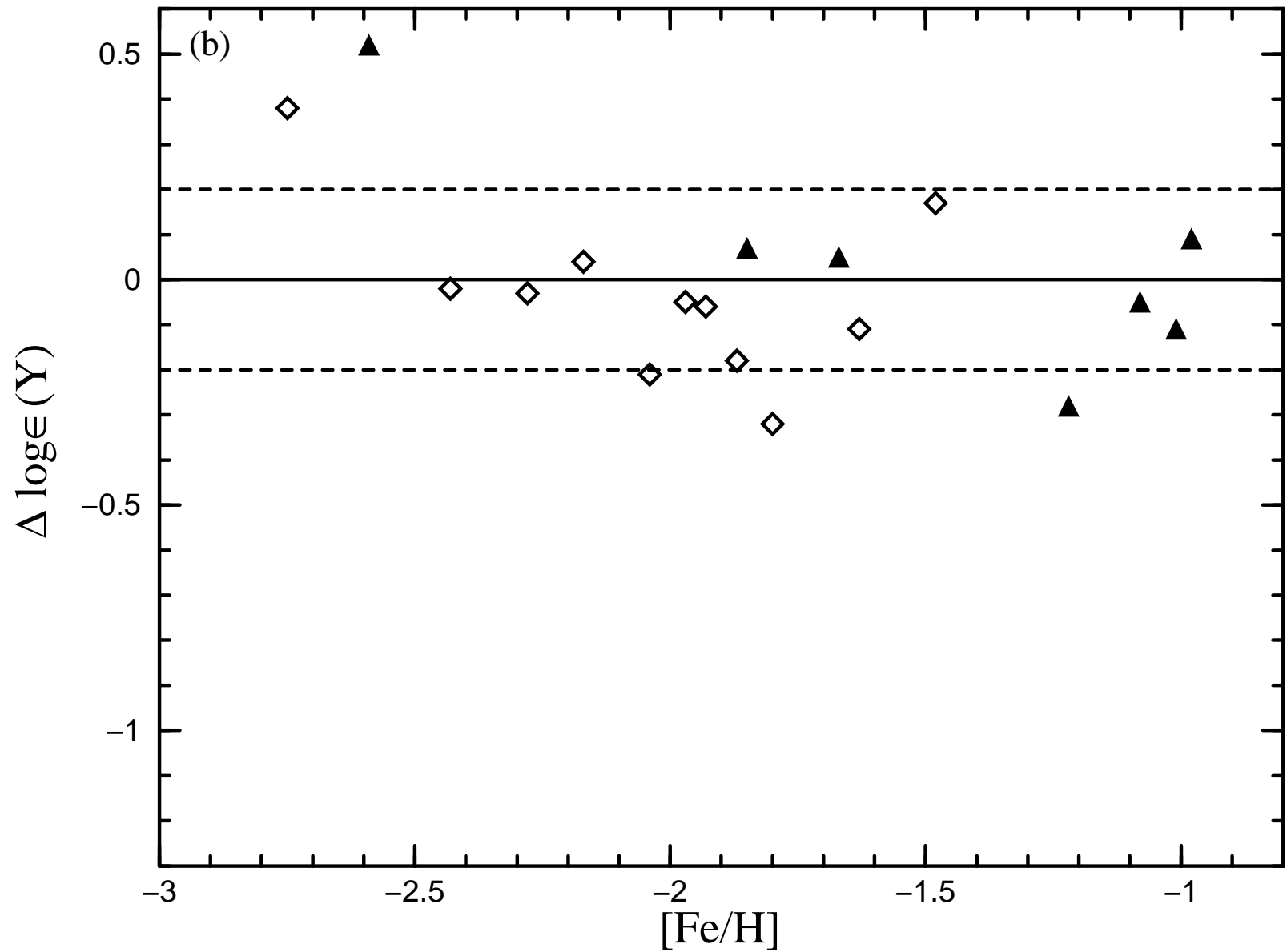
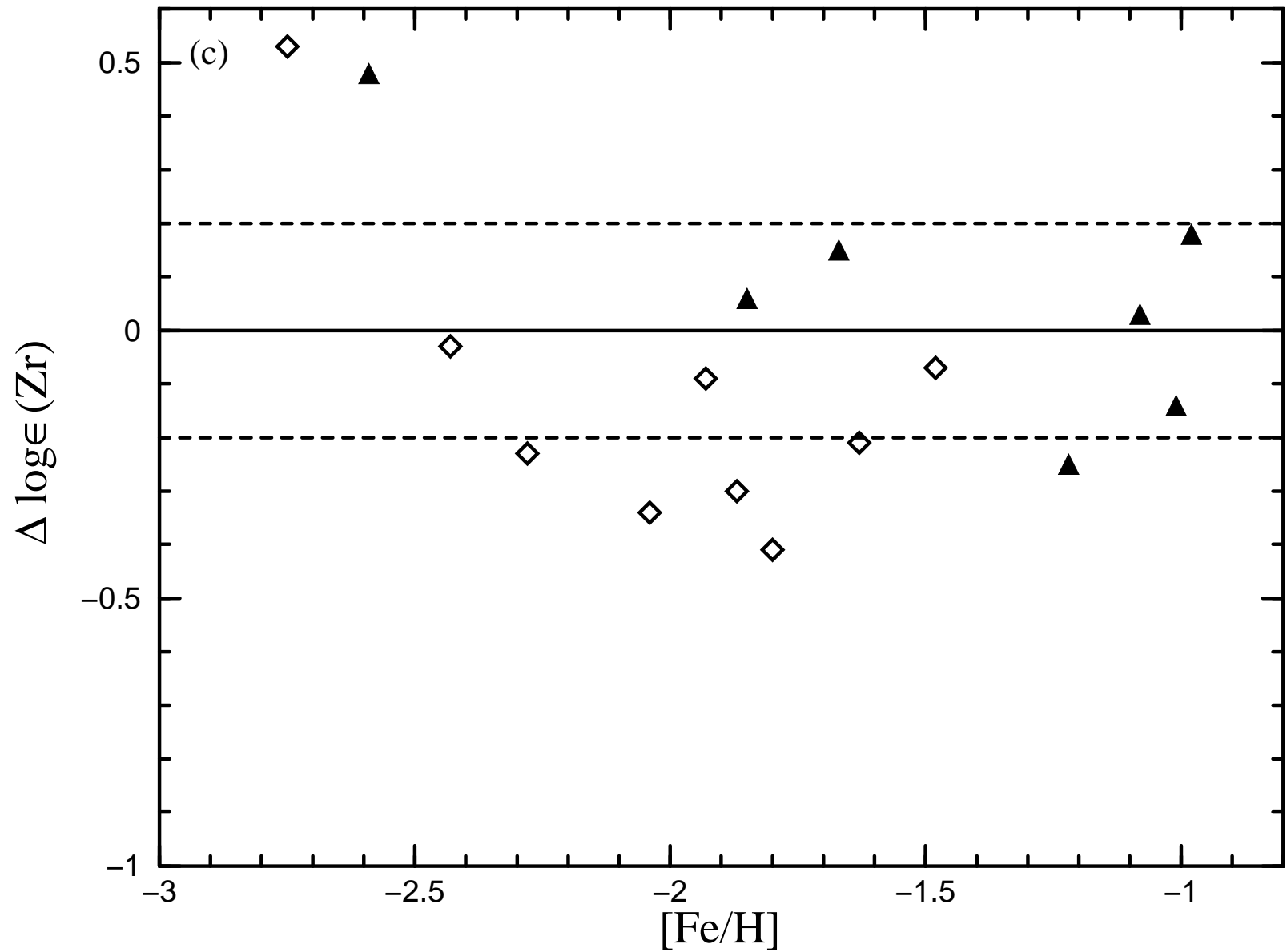
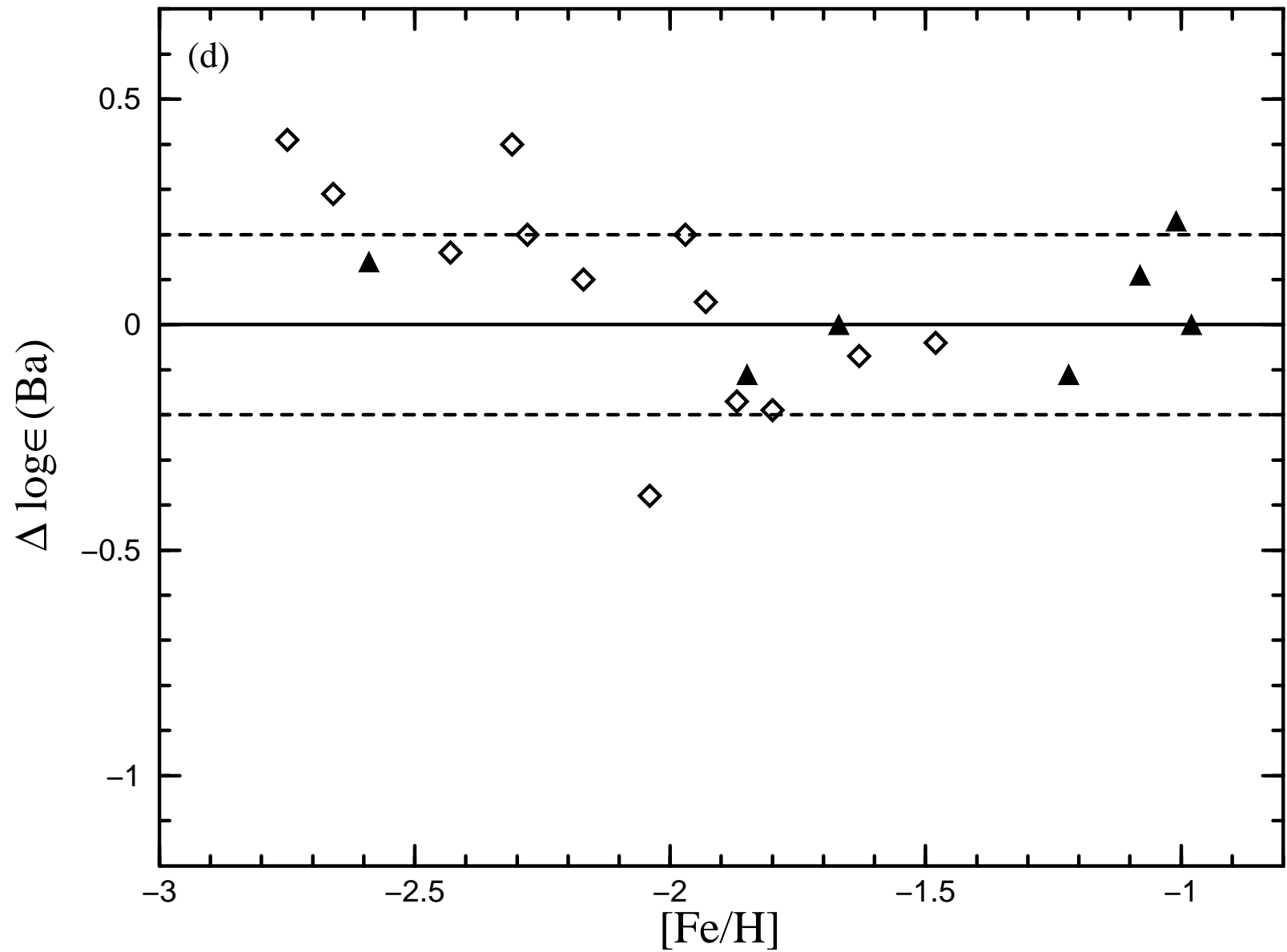
\end{document}